\documentclass[pdflatex,sn-mathphys-num]{sn-jnl}% Math and Physical Sciences Numbered Reference Style
\usepackage{graphicx}%
\usepackage{multirow}%
\usepackage{amsmath,amssymb,amsfonts}%
\usepackage{amsthm}%
\usepackage{mathrsfs}%
\usepackage[title]{appendix}%
\usepackage[dvipsnames]{xcolor}
\usepackage{textcomp}%
\usepackage{manyfoot}%
\usepackage{booktabs}%
\usepackage{algorithm}%
\usepackage{algorithmicx}%
\usepackage{algpseudocode}%
\usepackage{listings}%
\usepackage{natbib} 
\usepackage{times} 
\usepackage{amssymb,amsmath}
\usepackage{csquotes}
\usepackage{cancel}
\usepackage{dcolumn}% Align table columns on decimal point
\usepackage{bm}% bold math
\usepackage{widetext}
\usepackage{hyperref}
\theoremstyle{thmstyleone}%
\theoremstyle{thmstyletwo}%

\theoremstyle{thmstylethree}%

\renewcommand{\vec}[1]{\mathbf{#1}}
\newcommand{\vecg}[1]{\boldsymbol{#1}}
\newcommand{\tens}[1]{\mathbf{\underline{#1}}}

\begin{document}

\title[The effect of self-induced Marangoni flow on polar-nematic waves in active-matter systems]{The effect of self-induced Marangoni flow on polar-nematic waves in active-matter systems}


\author*[1]{\fnm{Andrey} \sur{Pototsky}}\email{apototskyy@swin.edu.au}

\author[2,3,4]{\fnm{Uwe} \sur{Thiele}}\email{u.thiele@uni-muenster.de}

\affil*[1]{\orgdiv{Department of Mathematics}, \orgname{Swinburne University of Technology}, \orgaddress{\street{Street}, \city{Hawthorn}, \postcode{3122}, \state{Victoria}, \country{Australia}}}

\affil[2]{\orgdiv{Institute for Theoretical Physics}, \orgname{University of M\"unster}, \orgaddress{\street{Wilhelm-Klemm-Str. 9}, \city{M\"unster}, \postcode{48149}, \country{Germany}}}
\affil[3]{\orgdiv{Center for Nonlinear Science (CeNoS)}, \orgname{University of M\"unster}, \orgaddress{\street{Corrensstr.\ 2,}, \city{M\"unster}, \postcode{48149}, \country{Germany}}}
\affil[4]{\orgdiv{Center for Multiscale Theory and Computation (CMTC)}, \orgname{University of M\"unster}, \orgaddress{\street{Corrensstr.\ 40}, \city{M\"unster}, \postcode{48149}, \country{Germany}}}

%%==================================%%Analysing

\abstract{We study the formation of  propagating large-scale density waves of mixed polar-nematic symmetry in a colony of self-propelled agents that are bound to move along the planar surface of a thin viscous film. The agents act as an insoluble surfactant, i.e.\ the surface tension of the liquid depends on their density. Therefore, density gradients generate a Marangoni flow.  
 We demonstrate that for active matter in the form of self-propelled surfactants with local (nematic) aligning interactions such a Marangoni flow nontrivially influences the propagation of the density waves. Upon gradually increasing the Marangoni parameter, which characterises the relative strength of the Marangoni flow as compared to the self-propulsion speed, the density waves broaden while their speed may either increase or decrease depending on wavelength and overall mean density. A further increase of the Marangoni parameter eventually results in the disappearance of the density waves. This may occur either discontinuously at finite wave amplitude via a saddle-node bifurcation or continuously with vanishing wave amplitude at a wave bifurcation, i.e.\ a finite-wavelength Hopf bifurcation.
 }




\maketitle

\section{Introduction}\label{intro}

Meso-scale propagating density fronts and wavy patterns are known to spontaneously arise in systems of active matter, i.e.\ of self-propelled interacting agents (for comprehensive reviews, see \cite{Ramaswamy2010,Marchetti2013,Baer2020}). These patterns emerge across a broad range of spatial scales, e.g., flocks of birds, fish, and insects on the meter-scale \cite{Vicsek1995,ToTu1998pre,ToTR2005ap,SoTa2013prl,SoCT2015prl,Couzin2005}, auto-chemotactic suspension of microorganisms \cite{Lushi2018} and density waves 
%of proteins on cell membranes \cite{Loose2008} or
 in actin motility assays \cite{Schaller2010} on the micrometer scale. Driven permanently out of equilibrium by the agents’ motility, resulting self-organised complex spatio-temporal dynamic regimes may be indefinitely sustained by chemical fuel.
%\ttuwe{are the proteins really self-propelled? We have to be careful not to mix active matter and standard RD systems as this would weaken the later argument.}

To model the characteristics of ``dry'' systems, where the motion of the fluid matrix supporting the self-propelled agents is not considered, the density distribution of the agents is described by the Boltzmann equation. It accounts for the microscopic rules governing binary collisions and random reorientations of the agents \cite{Bertin2009,Peshkov12,Denk20}. In the small-velocity regime, truncated forms of the Boltzmann equation can be derived by assuming dominant polar \cite{Bertin2009} or polar and nematic order parameter fields \cite{Peshkov12,Denk20}. Analysing the resulting set of closed hydrodynamic equations for these fields reveals that the observed spatio-temporal patterns in active matter originate in spontaneous symmetry breaking that occurs when the agent density surpasses a critical threshold \cite{Denk20}.

The situation and resulting models are more complex for ``wet'' systems, defined by a two-way coupling of the motion of the self-propelled agents and the surrounding fluid medium. In this scenario, the Boltzmann equation for the agents' density distribution is coupled to the Navier-Stokes equations that govern fluid motion. Here, we concentrate on a two-dimensional (2D) colony of self-propelled agents bound to move along the planar surface of a thin viscous film of liquid on a solid substrate. Then, the Boltzmann equation is extended by an advection term representing liquid flow at the film surface. Previously, we investigated the dynamics of a colony of noninteracting self-propelled agents on the surface of a liquid film with a deformable interface, particularly focusing on self-induced Marangoni flow \cite{PTS14,PTS16}. Such flows result from surface tension gradients that are due to concentration gradients of the agents.

When the average orientation of the self-propelled agents has a component orthogonal to the free surface, their motile force exerts an excess pressure similar to the hydrostatic pressure for a pending liquid film \cite{PTS14}. When the excess pressure dominates the stabilising surface tension, a Rayleigh-Taylor instability occurs \cite{Oron97,MPBT2005pf}. It is a long-wave stationary instability, i.e.\ its onset occurs at zero wavenumber and perturbations grow monotonically. In the classification of Ref.~\cite{FrTh2023prl} this corresponds to a Cahn-Hilliard type. In a passive system (i.e.\ one that approaches thermodynamic equilibrium) such an instability results in the formation of drops that slowly coarsen. However, the situation changes if also a component of motile force exist that is parallel to the free surface. This results in the generation of intricate self-propelled dynamic density patterns and is studied in Ref.~\cite{PTS16} by means of coupled thin-film and Smoluchowski equations. Note that in the context of the dynamics of an active Brownian particle, the Smoluchowski equation is essentially identical to the Boltzmann equation with an additional condition for the translational diffusivity constant which is linked to the temperature via the fluctuation-dissipation theorem \cite{MARCONI2008}.

In many practical scenarios, however, the orientation of the agents is always parallel to the liquid surface, which remains undeformed or weakly deformed at all times. An example of  an active-matter system that meets these conditions is a two-dimensional bacterial bath, formed by stretching a dilute bacterial suspension on a wire mesh to create a free liquid film with a thickness of several micrometers \cite{Wu2000,Sokolov2007,Sokolov2010}. In such cases, an isotropic disordered state may be unstable due to local interactions among the agents, driving the system towards spatio-temporal coherent dynamics and chaos.

  Here, we consider a model system of swimmers moving parallel to the non-deformable free surface of a liquid film on a solid substrate, and study the dynamics of the relevant order parameter fields. In this way we explore how the advective flow of the liquid matrix, induced by surface tension gradients affects the propagation of density waves. We expand a macroscopic continuum approach developed for dry active systems with local nematic alignment \cite{Peshkov12,Denk20}. In particular, we consider the wet active system that results from the incorporation of self-induced Marangoni flow into the hydrodynamic mean-field equations for the dominant polar and nematic order parameter fields.

  %Our subsequent bifurcation analysis reveals a strong similarity to the influence of self-induced Marangoni flow on travelling reaction-diffusion waves confined in a liquid drop or film \cite{Rongy2006,Rongy2007}. There, the reaction is fuelled by an exothermic autocatalytic chemical reaction, such as the Belousov-Zhabotinsky or the iodate-arsenous acid reaction, and all reactants are non-motile. 
    % \ttuwe{we need to be more specific what is similar!}
%
%    Here, we use the convention that an active-matter system is formed by particles (agents) with intrinsic motility, while a passive-matter system is composed of passive particles that do not exhibit a sustained motion when separated from the system. } \ttuwe{need to discuss the red part, as I would like to write this in a way consistent to my other works.} {\color{red} a. \tt please feel free to change if necessary according to your earlier works. }

Specifically,  we demonstrate that upon gradually increasing the Marangoni parameter, which characterises the relative strength of the Marangoni flow as compared to the self-propulsion speed of the agents, density waves broaden while their speed may either increase or decrease depending on wavelength and on overall mean density. This result echos the influence of self-induced Marangoni flow on travelling reaction-diffusion waves confined in a liquid drop or film \cite{Rongy2006,Rongy2007}. There, the reaction is fuelled by an exothermic autocatalytic chemical reaction, such as the Belousov-Zhabotinsky or the iodate-arsenous acid reaction, and all reactants are non-motile. Similar to our observations, the inclusion of the Marangoni flow in autocatalytic chemical reaction systems leads to a change in propagation speed and front deformation.
We also demonstrate two different bifurcation scenarios that result in the destruction of the waves at a relatively strong Marangoni flow. In the first scenario, a gradual increase in the Marangoni parameter results in the approach and ensuing collision of branches of stable und unstable finite-amplitude waves. The corresponding saddle-node bifurcation results in a discontinuous disappearance of the density waves. In the second scenario, a gradual increase in the the Marangoni parameter results in a gradual decrease of the wave amplitude until it eventually vanishes in a continuous transition via a wave bifurcation, i.e.\ a finite-wavelength Hopf bifurcation, of a uniform state. 

%\ttuwe{I now suppressed the active/passive discussion, by only using ``active matter'' or ``self-propelled agents'', and not using ``passive'', as also an RD system is an active system, but does not involve active matter (in the sense of ``self-propelled''. But as we are not writing a review it is not relevant, as long as we do not use ''passive''. The latter should only be used for systems that settle into a thermodynamic equilibrium.}

\section{Governing equations }
\label{model}
\begin{figure*}[ht]
  \centering
  \includegraphics[width=0.9\textwidth]{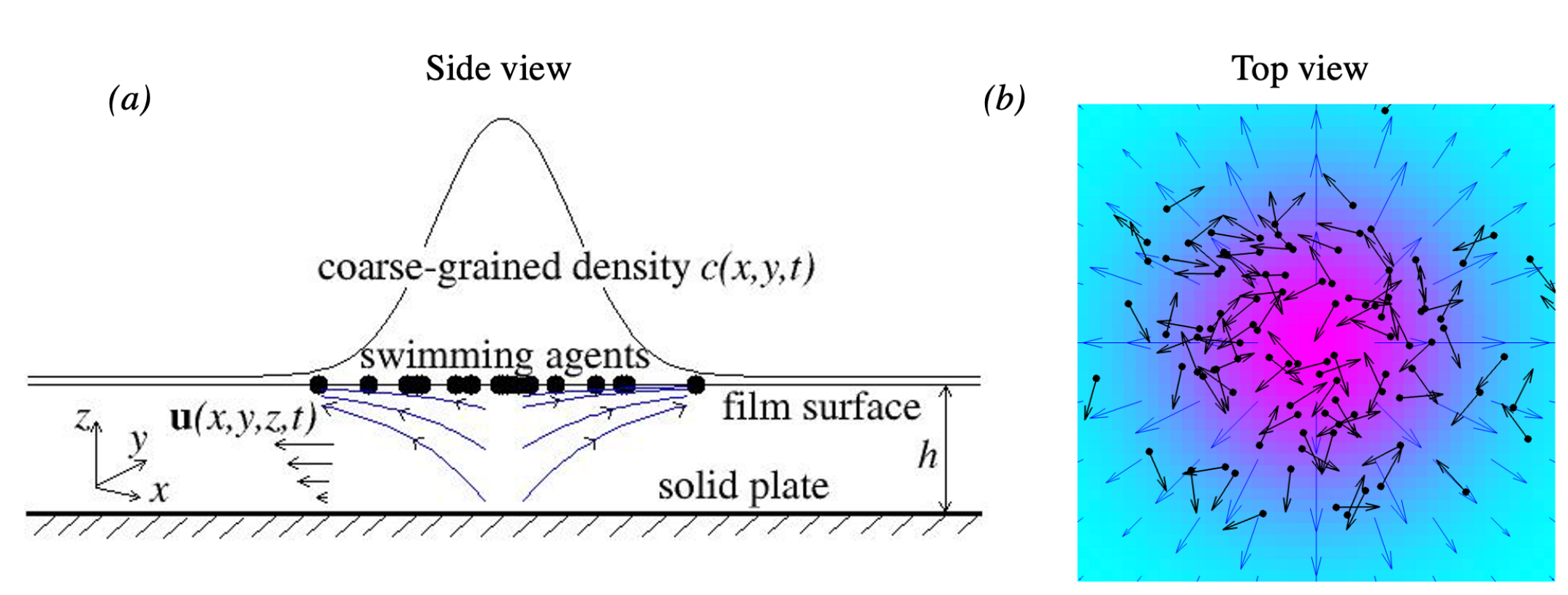}
   \caption{\label{Fsketch} Schematic representation of the system in (a) side view and (b) top view.  Swimming agents represented by filled circles move along a flat liquid-air interface of a thin liquid film of uniform thickness $h$. On a coarse-grained level, swimmers of concentration $c(x,y,t)$ act as a surfactant, inducing Marangoni flow with velocity $\vec{u}(x,y,z,t)$ directed away from the region with higher density. The agents are advected by the surface flow field $\vec{U}(x,y,t)=\vec{u}(x,y,z=h,t)$, shown by blue arrows in (b). The orientations of individual agents are indicated by black arrows in (b) while the coarse-grained particle density is color-coded.}
   %\ttuwe{in sketch: add $c(x,y,t)$ below ``coarse-grained density''; make $u$ bold face}}
\end{figure*}
Consider a colony of self-propelled agents that are bound to move along a planar surface of a thin liquid carrier film of uniform and constant thickness $h$ and dynamic viscosity $\mu$ that is supported by a solid bottom plate, as schematically shown in Fig.\ref{Fsketch}. The orientation of all agents remains parallel to the film surface at all times.  Any two agents interact with each other via local binary collision rules. Thereby, any two colliding agents will align along their average pre-collision orientation when colliding at an acute angle and anti-align when colliding at an obtuse angle. Here, we adopt the kinetic Boltzmann approach developed for dry active matter in the absence of a fluid matrix \cite{Peshkov12,Denk20}. 

To describe the interaction with the liquid, we neglect the size of the agents and assume that the main contribution to their motion is due to the advection by the liquid along the surface of the carrier film with the local velocity $\vec{U}=(U_x,U_y)=\vec{u}(x,y,z=h,t)$, where $\vec{u}(x,y,z,t)$ is the horizontal flow field in the bulk.  We emphasise that the motion of solid particles of finite size trapped at an interface between two fluids is a complex hydrodynamic problem that (in the limit of small Reynolds number) involves the solution of the Stokes equation supplemented with no-slip boundary conditions at the particle surface and the continuity conditions for the flow field involving stresses at the interfaces as well as wetting conditions. For a comprehensive review of the motion of passive colloidal particles at liquid interfaces see e.g.\ \cite{Anderson89}.  The complete hydrodynamic description becomes even more involved for self-propelled particles moving near penetrable liquid-liquid interfaces and can only be resolved numerically \cite{Feng23}. Here, we treat the particles as point-like implying that they are advected with the velocity $\vec{U}$.

 On the continuum level, the ensemble of self-propelled agents at the film surface is described by the one-particle probability density distribution $\rho(\vec{r},\theta,t)$ that depends on the position in the film surface plane $\vec{r}=(x,y)$ (that is parallel to the solid substrate) and the orientation $\theta$ of the agents. Its time evolution is described by the kinetic Boltzmann equation
\begin{eqnarray}
\label{eq1}
\partial_t \rho+\vec{\nabla}\cdot [(\vec{U}+v_0 \vec{e}(\theta))\rho]=d\vec{\nabla}^2 \rho + I_{\text{rot}}+ I_{\text{coll}},
\end{eqnarray}
where $v_0$ is the self-propulsion speed, $\vec{e}(\theta)=(\cos(\theta),\sin(\theta))$ is a unit orientation vector, $d$ is the translational diffusivity, $I_{\text{rot}}$ describes rotational diffusion and $I_{\text{coll}}$ is the collision term that describes the alignment of the agents.  If the one-particle density distribution $\rho(\vec{r},\theta,t)$ is normalised in such a way that $\int_A d\vec{r}\int_0^{2\pi}d\theta \rho(\vec{r},\theta,t)=N$, where $N$ is the total number of agents for a liquid film of surface area $A$, then their local number density per surface area is $c(\vec{r},t)=\int_0^{2\pi}d\theta\,\rho(\vec{r},\theta,t)$. Then $c_0=A^{-1}\int_A c(\vec{r},t)\,d\vec{r}=NA^{-1}$ is the mean number density.

  The rotational diffusion term $I_{\text{rot}}$  in equation~(\ref{eq1}) depends on the one-particle density $\rho(\vec{r},\theta,t)$, while for binary collisions, the collision term $I_{\text{coll}}$ depends on the two-particle density function $\rho_2(\vec{r}_1,\theta_1,\vec{r}_2,\theta_2,t)$.  However, in the mean field approximation one truncates the resulting hierarchy of equations assuming the closure relation $\rho_2(\vec{r}_1,\theta_1,\vec{r}_2,\theta_2,t)=\rho(\vec{r}_1,\theta_1,t)\rho(\vec{r}_2,\theta_2,t)$  \cite{Bertin2006}.
     This allows one to represent  $I_{\text{coll}}$ as a quadratic functional of $\rho$,  as described in Refs.~\cite{Peshkov12,Denk20}.

  Equation~(\ref{eq1}) must be supplemented with an equation for the advection flow field $\vec{U}(\vec{r},t)$. Similar to Ref.~\cite{PTS14,PTS16} we assume that in the presence of the carrier liquid the self-propelled agents  act as a surfactant, i.e.\ due to the solutal Marangoni effect, gradients in their density induce a Marangoni flow at the film surface. Note that translational diffusion and linear Marangoni effect both result from entropic contributions of an underlying energy functional as shown for insoluble surfactants in Ref.~\cite{ThAP2012pf}.  
  %A more realistic model may also include the concentration of chemicals secreted by the agents for example as the result of the metabolic activity (\color{red}{of self-propelled bacteria \cite{Sokolov2007}.  Such chemicals are produced} at a local rate which is proportional to the local concentration of agents and decompose via the reaction of solidification. 
  %\ttuwe{do you want to include this? It raises more questions as it seems to imply we somehow model this.}
 % However, in the limit of fast reactions and slow advection-diffusion, the local concentration of the chemicals is proportional to the local concentration of the agents. Under this assumption, the concentration of chemicals is locally enslaved to the concentration of the agents. 
  %\ttuwe{I find this discussion difficult as it somehow says marangoni is not due to swimmer concentration but due to something else.}

The flow field generated by a non-uniform concentration field $c$ can be found by solving the Stokes equation in the fluid supplemented with the Marangoni stress boundary condition at the film surface $\mu \partial_z \vec{u}(x,y,z,t)\big\vert_{z=h}=- \frac{d\sigma}{dc} \vec{ \nabla}c$,  where $\sigma(c)$ is the concentration-dependent surface tension.
  
  % \ttuwe{We can not use $\vec{U}$ for the surface velocity AND the full velocity field! Better introduce ($\vec{u}(\vec{r},z,t)$ and $\vec{U}=\vec{u}(\vec{r},h,t)$ where $\vec{u}=(u,v)^T$).}  
  This problem has been solved in Ref.~\cite{Thess97} by assuming that the deformation of the film surface can be neglected, which is valid at small capillary numbers $\mu U/\sigma_0\ll 1$, where $\sigma_0$ is the surface tension of the bare liquid in the absence of surfactant. However, at large capillary number a complete description of the dynamics of a wet system would necessarily include the dynamics of film surface deformations. This will be addressed in a future study.

 In particular, it was shown in Ref.~\cite{Thess97} that for thick liquid layers of uniform thickness the surface velocity $\vec{U}$ induced by the Marangoni effect depends nonlocally on gradients of the surfactant density $c$ and can be represented as
\begin{eqnarray}
\label{eq2}
\vec{U}=-\frac{E h }{4\pi^2 c_0 \mu}\int d^2k \frac{i\vec{k} \hat{c}(\sinh^2(|\vec{k}|h)-(|\vec{k}|h)^2)}{|\vec{k}|h\sinh(2|\vec{k}|h)-2(|\vec{k}|h)^2}e^{i\vec{k}\cdot \vec{r}}.
\end{eqnarray}
Here, $E=-c_0(d\sigma(c)/dc)\big\vert_{c=c_0}>0$ is the Marangoni elasticity of the film surface that encodes the strength of change of the surface tension $\sigma(c)$ with $c$. Further, $\vec{k}=(k_x,k_y)$ is a 2D wave number and $ \hat{c}(\vec{k},t)=4\pi^2 \int d^2r\, c(\vec{r},t)e^{-i\vec{k}\cdot \vec{r}}$ is the Fourier transform of $c$.

In long-wave approximation \cite{Oron97,Thie2007chapter,CrMa2009rmp},  valid when the thickness of the liquid film is small as compared to the characteristic horizontal length scale of the variation in surfactant density, i.e.\ $h\ll 2\pi/|\vec{k}|$, Eq.~\eqref{eq2} reduces to the local relationship \cite{Oron97}
 \begin{eqnarray}
\label{eq4}
\vec{U}(\vec{r},t)=-\frac{Eh}{4 c_0 \mu} \vec{\nabla} c(\vec{r},t).
\end{eqnarray}
It is interesting to observe that the functional dependence of $\vec{U}$ on $c$ in equation~(\ref{eq4}) is identical to the one found for the velocity of a spherical solid colloidal particle transported at an interface by diffusiophoresis \cite{Anderson89}.
%Recently, the local approximation Eq.\,(\ref{eq4}) has been used to study the dynamics of a two-dimensional colony of Marangoni surfers \cite{Gouiller21}.

The uniform disordered state  $\rho(\vec{r},\theta)=c_0/(2\pi)$ may become unstable to spatial and orientational perturbations, leading to a transition toward dynamic inhomogeneous ordered states including steady clusters or bands, and propagating waves as it was demonstrated previously for dry system \cite{Bertin2009,Peshkov12,Denk20}. One of the key observations derived from the numerical analysis of the Boltzmann equation~(\ref{eq1}) is that despite the short-range interactions due to collisions, which typically occur when the distance between the agents is comparable with their size, the global dynamics of the system exhibits a long-range order with a much larger characteristic scale.
This allows us to replace the Boltzmann equations for the one-particle density $\rho(x,y,\theta,t)$ with a set of coupled hydrodynamic equations for certain coarse-grained order parameter fields that do not depend on the orientation angle $\theta$. As such, the Fourier transform of the density in the angular space is used
%\ttuwe{Before starting with the order parameter field, the general gist should be explained: (i) have already truncated (1) via closure relation,  (ii) expand the remaining  one-particle density $\rho(\vec{r},\theta,t)$ in angular Fourier modes, (iii) truncate hierarchy of equations using scaling argument. At the moment there is quite a jump here.}

  The orientation order on a large scale is well described by the first few order parameter fields
\begin{equation}
\label{eqorderParameters}
\rho_q(\vec{r},t)=\int_0^{2\pi} e^{iq\theta}\rho(\vec{r},\theta,t)\,d\theta.
\end{equation}
In fact, for polar or polar-nematic alignment between the agents during collisions, the orientation order is well approximated by the first four dominant order parameters $\rho_{1,2,3,4}$ with all other order fields assumed to be negligibly small, i.e. $\rho_{q>4}=0$.

Our goal is to follow the closure scheme outlined in \cite{Peshkov12,Denk20} to derive a closed set of coarse-grained hydrodynamic equations for the density $c$ and the dominant order parameters $\rho_{1,2}$ for a wet system by including the advection velocity $\vec{U}$. Thus, we also assume the following scaling and truncation
\begin{align}
  \begin{split}
&|\vec{U}(c)|\sim v_0,~~\partial_{t,x,y}\sim \epsilon,~~c-c_0\sim \epsilon,\\
&\rho_{1,2}\sim \epsilon,~~\rho_{3,4}\sim \epsilon^2,~~\rho_{q>4}=0
\end{split}
\label{scale}
\end{align}
where $\epsilon\ll 1$ is the dimensionless super-criticality, i.e.\ the distance from the onset of the order-disorder transition. From the physical point of view, the above scaling implies a slow dynamics with a long-wave spatial orientation order dominated by polar $\rho_1$ and nematic $\rho_2$ order parameter fields. The scaling~\eqref{scale} automatically restricts the applicability of the hydrodynamic approximation to relatively small magnitudes and slow variations of $\rho_{1}$ and $\rho_{2}$.  In consequence, similar to other weakly nonlinear approaches, the approximation is valid only in the vicinity of the order-disorder transition where the disordered state $c=c_0$ becomes unstable.

The closed form of the dynamic equations for the density $c$ and the complex-valued polar and nematic order parameter fields $\rho_1$ and $\rho_2$ is obtained in several steps as detailed in Appendix~\ref{appendix}. First, the Boltzmann equation is Fourier-transformed in the angular space and truncated at $q=4$, i.e. all order fields with $q>4$ are identically zero. Next, using scaling arguments~(\ref{scale}), the fields $\rho_{3,4}$ are expressed in terms of $\rho_{1,2}$. Finally, after substituting $\rho_{3,4}$ into the equations for $c$ and $\rho_{1,2}$, the closed set of hydrodynamic equations is obtained as
 %(\color{red}{when using the dynamic equations for $\rho_{3,4}$ resulting when writing the Boltzmann equation \eqref{eq1} in terms of the first angular Fourier modes to express $\rho_{3,4}$ in terms of $\rho_{1,2}$. This is detailed in Appendix~\ref{appendix}.}
%
\begin{eqnarray}
\partial_t c + \vec{\nabla}\cdot (\vec{U}c)&=&-\frac{v_0}{2}\left(\partial \rho_1^* + \partial^* \rho_1 \right)+d\vec{\nabla}^2 c\nonumber\\
\partial_t \rho_1+\vec{\nabla}\cdot (\vec{U}\rho_1)&=&-(\alpha_0+c\alpha_1)\rho_1+\alpha_2\rho_1^*\rho_2
-\alpha_3 |\rho_2|^2\rho_1\nonumber\\
&-&\frac{v_0}{2}(\partial c + \partial ^* \rho_2)+\gamma_1 \rho_2^*\partial \rho_2+d\vec{\nabla}^2 \rho_1 \label{eq5}\\
\partial_t \rho_2+\vec{\nabla}\cdot (\vec{U}\rho_2)&=&(-\beta_0+c\beta_1)\rho_2+\beta_2\rho_1^2-\beta_3 |\rho_2|^2\rho_2\nonumber\\
&-&\beta_4|\rho_1|^2\rho_2-\frac{v_0}{2}\partial \rho_1 +\gamma_2 |\partial|^2 \rho_2 \nonumber\\
&-&\gamma_3 \rho_1^* \partial\rho_2 -\gamma_4  \partial^*(\rho_1\rho_2)+d\vec{\nabla}^2 \rho_2.\nonumber
\end{eqnarray}
Here, $\partial =\partial_x + i\partial_y$ and $\vec{\nabla}=(\partial_x,\partial_y)^T$ while the star indicates the complex conjugate. The system \eqref{eq5} is closed by Eq.~\eqref{eq2} [or Eq.~\eqref{eq4} in the long-wave limit] that relates $\vec{U}$ to the number density $c$. Note that we have retained translational diffusion with diffusivity $d$  to ensure numerical stability and consistency with the Marangoni term. For ease of comparison, system~\eqref{eq5} is nondimensionalised using the scaling introduced in Ref.~\cite{Denk20} for a dry system, i.e.\ when $\vec{U}=0$. Namely, time is scaled with the characteristic duration of the ballistic flight path of the agents $\tau=r^{-1}$, where $r$ is the rate of random orientation changes of the agents. The spatial coordinates are scaled with $v_0\tau$, and the one-particle density distribution $\rho$ is measured in units of $r/Dv_0$, where $D$ is the size of the agent particles.  In dimensionless units, the dynamics of a wet system in the long-wave limit is described by  Eqs.~\eqref{eq5} with $v_0=1$ and $\vec{U}=-\text{Ma}\vec{\nabla}c$ where $\text{Ma}=E h/(4 v_0^2\mu \tau)$ represents the Marangoni parameter that controls the relative strength of the Marangoni flow as compared to the self-propulsion speed.

The coefficients $\alpha_i$ and $\beta_i$ in Eqs.~\eqref{eq5} determine the properties of the spatially uniform states, while the $\gamma_i$ are responsible for the coupling between the gradients of polar and nematic order parameter fields.  

As outlined in Appendix~\ref{appendix}, the closure relations for $\rho_{3,4}$ and, therefore,  the coefficients in Eqs.~\eqref{eq5} depend on the average density $c_0$, as well as on the microscopic parameters such as the statistics of the rotational diffusion, the effective interaction distance and the alignment rules associated with binary collisions. The behaviour of the system will therefore depend on changes in these microscopic parameters even if the average coarse-grained density is kept fixed.  However, since our aim is to understand the effect of the self-induced Marangoni flow on the dynamics of the system on the coarse-grained level, we apply a semi-phenomenological approach as proposed in Ref.~\cite{Denk20}, i.e.\ we derive the governing equations by coarse graining but then pursue a parametric study on the continuum level. Accordingly, the obtained hydrodynamic equations \eqref{eq5} are studied for selected physically meaningful control parameters while all other parameters remain fixed. In particular, our main focus is the effect of the Marangoni number $\text{Ma}$ and of the mean  density $c_0$. Note that $\text{Ma}$ can be varied independently at fixed density for a given set of the microscopic parameters.

Following Ref.~\cite{Denk20}, we choose the coefficients $\alpha_i$ and $\beta_i$ to correspond to a system that is known to feature several transitions in absence of Marangoni flow. When the mean number density $c_0$ is varied at fixed values of the coefficients $\alpha_i$, $\beta_i$ and $\gamma_i$, transitions occur between a uniform disordered state and a uniform ordered nematic state and as well between the latter and a uniform ordered polar-nematic state of mixed symmetry. The parameter values employed in the following are
\begin{eqnarray}
\label{coeff}
\alpha_0&=&0.011,~\alpha_1=0.44,~\alpha_2=1.5,~\alpha_3=4.4,\nonumber\\
\beta_0&=&0.044,~\beta_1=0.59,~\beta_2=-0.16,~\beta_3=6.6,~\beta_4=-1.0,\nonumber\\
\gamma_1&=&1,~\gamma_2=0.85,~\gamma_3=0.23,~\gamma_4=3.7.
\end{eqnarray}
 Coefficients $\alpha_i$, $\beta_i$ and $\gamma_i$ can be traced back to the microscopic parameters and the average density via the relations given in Appendix~\ref{appendix}. In fact, the values used here correspond to $c_0=0.08$ for the choice of the collision rules used in Ref.~\cite{Denk20}.  The translational diffusivity is varied between $d=10^{-2}$ and $d=10^{-1}$.

\section{Marangoni flow-induced change in polar-nematic waves}
\label{sec:2d}
As shown in Ref.~\cite{Denk20}, the dry system, i.e.\ without the presence of Marangoni flow, develops polar-nematic travelling waves for mean densities above $c_0=\beta_0/\beta_1=0.074$. Before embarking on a detailed analysis, we first demonstrate that self-induced Marangoni flow influences the propagation of these density waves. We use the long-wave variant  [Eq.~\eqref{eq4}] to calculate the Marangoni flow \cite{note} and numerically solve Eqs.~\eqref{eq5} in 2D using central differences for spatial derivatives and a semi-implicit second-order Heun method for time integration. As initial condition, we use a uniform disordered state with small-amplitude random initial perturbations with a slight polar and nematic bias in the $x$-direction. Eqs.~\eqref{eq5} are numerically integrated over a sufficiently long time interval of $5000$ units at $\text{Ma}=0$ to allow the system to reach a steady state. Then the average wave number associated with the Fourier-transformed density field $\int |\vec{k}| \hat{c}(\vec{k},t)\,d\vec{k}$ fluctuates about a constant value and does not show any trend as a function of time. The density field $c$ recorded over the subsequent time interval $\Delta t=1000$ is used to extract the statistics of the speed of the waves (not shown). The dynamics of the system at $\text{Ma}>0$ is obtained using naive parameter continuation. Namely, after the system is integrated for $\Delta t=6000$, the Marangoni parameter $\text{Ma}$ is abruptly incremented by $10$ and Eqs.~\eqref{eq5} are further integrated without setting a new initial condition.

\begin{figure*}[ht]
  \centering
  \includegraphics[width=0.9\textwidth]{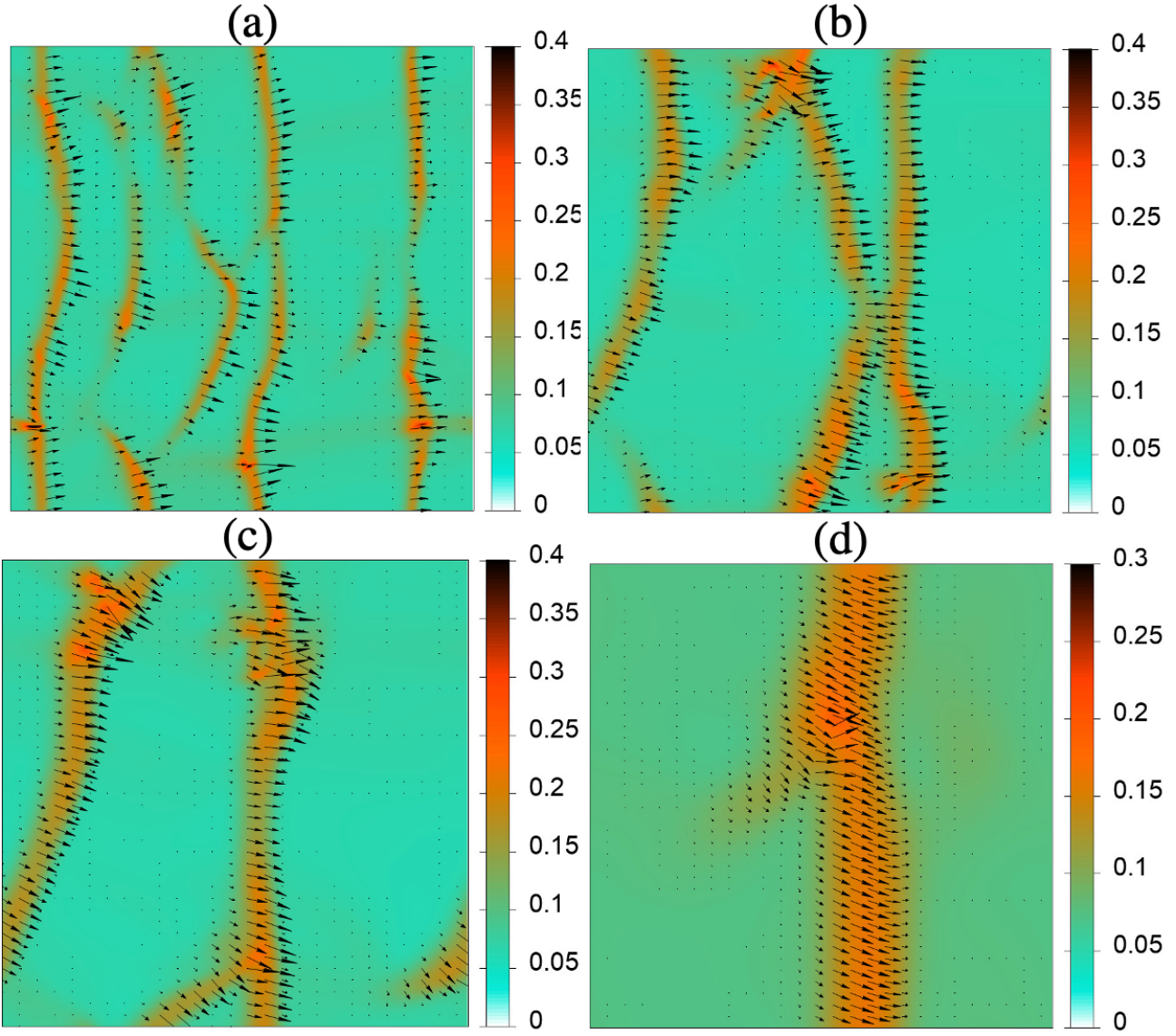}
   \caption{\label{F1} Snapshots of fully developed polar-nematic travelling waves in 2D for different values of the Marangoni parameter illustrate the effect of Marangoni flow. Shown are (a) $\text{Ma}=0$, (b) $\text{Ma}=10$, (c) $\text{Ma}=30$, and (d) $\text{Ma}=150$. Accompanying movies in the Supplementary Material further illustrate the wave dynamics. The mean density is $c_0=0.09$, the diffusivity is $d=0.1$, and a $800\times 800$ domain with periodic boundaries is used. The background colouring indicates the number density $c$ as indicated in the color bar. Black arrows indicate the vectorial polar order parameter field $(\text{Re}(\rho_1),\text{Im}(\rho_1))$.  No stable waves are found for $\text{Ma}>160$. All remaining parameters are given in the main text.}
\end{figure*}

Snapshots of fully developed travelling wave patterns are shown in Fig.~\ref{F1} for $c_0=0.09$ and $d=0.1$ in a square domain of size $800\times 800$ with periodic boundaries. These waves appear to be long-time stable and propagate without persistent coarsening or splitting. %The background color represents the number density field $c$, while arrows indicate polar order vector field with coordinates $(\text{Re}(\rho_1),\text{Im}(\rho_1))$. 
Inspecting Fig.~\ref{F1}, one discerns that the inclusion of Marangoni flow results in a broadening of the polar-nematic wave in the direction of propagation as well as a moderate increase in its average propagation speed that can be estimated from the movies in the Supplementary Material (SM). The dependence of the propagation speed of the waves on the Marangoni parameter $\text{Ma}$ is addressed in detail in the next section.  

An interesting feature of the travelling waves is their apparent stability with respect to coarsening, i.e.\ for any fixed value of $\text{Ma}$, the waves seemingly propagate without merging to form a broader wave of larger wavelength. Coarsening occurs only during the transient phase directly after an abrupt increase of $\text{Ma}$ when the system evolves into a new stationary state with a larger average wavelength. Thus, upon incrementally increasing the Marangoni parameter, the number of wave crests reduces from five in the dry system at $\text{Ma}=0$ to two at $\text{Ma}=30$ and finally only one at $\text{Ma}=150$. Further increasing the Marangoni parameter beyond a certain critical value ($\text{Ma}\gtrapprox 160$), the wave dissipates into a uniform nematic state.

Suppressed or completely arrested coarsening is a well-known phenomenon in active systems like matter driven by external forces, active matter (i.e.\ motile matter) and reaction-diffusion waves. Generally, there exist two physical mechanisms that underpin coarsening, namely, (i) the volume mode (also known as Ostwald ripening or mass transfer) and (ii) the translation mode (also known as coalescence where two or more clusters merge into one) \cite{Thie2007chapter,GORS2009ejam,PTA14}.
In active systems formed by non-motile agents, suppression and arrest of coarsening is observed in the presence of an external driving force studied, e.g., with convective Cahn-Hilliard models \cite{Zaks05,TALT2020n}, or is caused by nonequilibrium chemical potentials studied, e.g.\ with nonreciprocal Cahn-Hilliard models \cite{Frohoff2020}.

In systems of  active matter, i.e.\ composed of motile agents, self-propulsion leads to the formation of clusters of agents that do not merge to form a single dense phase \cite{Palacci2010,Palacci2014,Palacci2020}, As discussed in \cite{GONNELLA2015}, the exact physical mechanisms behind arrested coarsening in active matter systems are still unclear. However, on the phenomenological level, suppressed coarsening could be due to the non-variational nature of the evolution equations Eqs.~\eqref{eq5} which is inherent to all active matter systems. The question of the exact physical origin of the arrested coarsening requires a detailed stability analysis of various travelling wave states and will be addressed in future studies.
%\ttuwe{I think, in the final paragraph, we need either to weaken statements or go deeper into the literature. To my knowledge the standard motility-induced phase separation as described by a CH equation shows coarsening. But I would need to look more deeply into the literature to see which papers talk about coarsening and which do not.}
% 
\section{One-dimensional polar-nematic waves}
\label{1d}
 To gain a deeper understanding of the influence of Marangoni flow on the characteristics of the density waves, we next analyse a one-dimensional (1D) approximation. Namely, we assume that waves like the ones shown in Fig.~\ref{F1} are nearly translationally invariant in the direction orthogonal to the direction of propagation. In particular, we consider waves that travel in $x$-direction and are invariant in $y$-direction. This implies that all fields depend only on the $x$-coordinate, that the imaginary parts of the order parameter fields are zero $\text{Im}(\rho_1)=\text{Im}(\rho_2)=0$, i.e.\ Eqs.~\eqref{eq5} with \eqref{eq4} become
\begin{eqnarray}
\label{eq6}
&&\partial_t c -\text{Ma} \,\partial_x ( c\partial_x c)=- \partial_x \rho_1+d\partial_x^2 c\nonumber\\
&&\partial_t \rho_1-\text{Ma}\,\partial_x ( \rho_1 \partial_x c)=-(\alpha_0+c\alpha_1)\rho_1+\alpha_2\rho_1\rho_2\nonumber\\
&&-\alpha_3 \rho_2^2\rho_1-\frac{1}{2}(\partial_x c + \partial_x \rho_2)+\gamma_1 \rho_2\partial_x \rho_2+d\partial_x^2 \rho_1\nonumber\\
&&\partial_t \rho_2-\text{Ma}\,\partial_x (\rho_2\partial_x c)=(-\beta_0+c\beta_1)\rho_2+\beta_2\rho_1^2-\beta_3 \rho_2^3\\
&&-\beta_4\rho_1^2\rho_2-\frac{1}{2}\partial_x \rho_1 +(\gamma_2+d) \partial_x^2 \rho_2 -\gamma_3 \rho_1 \partial_x \rho_2 \nonumber\\
&&-\gamma_4  \partial_x(\rho_1\rho_2).\nonumber
\end{eqnarray}
Note that the system~\eqref{eq6} is invariant under the transformation $(t,x,c,\rho_1,\rho_2)\rightarrow (t,-x,c,-\rho_1,\rho_2)$, i.e.\ it is of mixed parity. In the following, we analyse the 1D model by determining uniform steady states and their linear stability as well as travelling wave states.

 \begin{figure}[ht]
     \centering   \includegraphics[width=0.99\columnwidth]{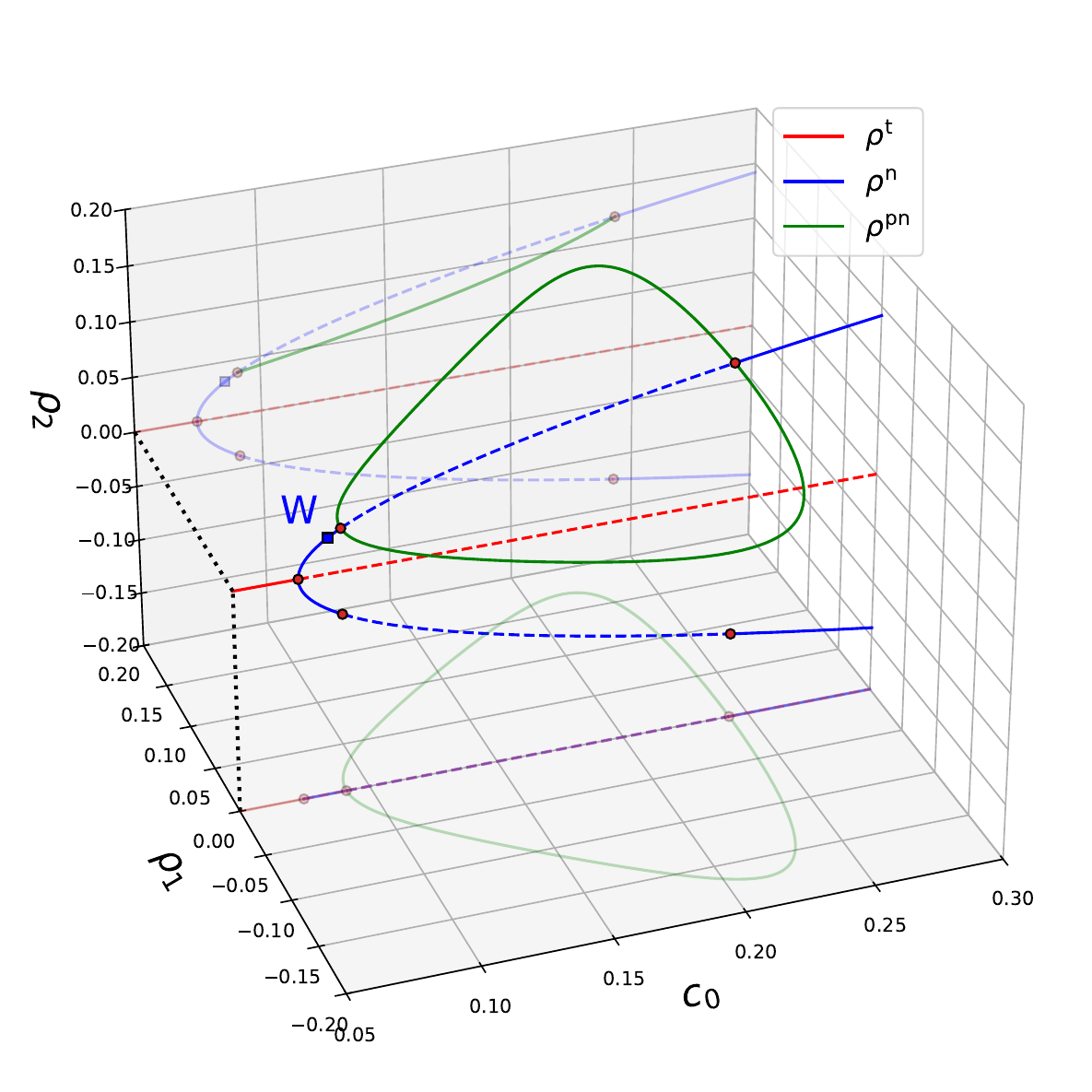}
        \caption{\label{F0}  Bifurcation diagram of the uniform steady states $(\rho_{1}^\mathrm{s}, \rho_{2}^\mathrm{s})$ as a function of the mean density $c_0$ in a 3D representation. Shown are the trivial state $\rho_{1,2}^{\mathrm{t}}=0$ (red line) from which the purely nematic state $\rho_{1,2}^\mathrm{n}$ [blue lines, Eq.~\eqref{nem}] emerge in a pitchfork bifurcation at $c_0=0.074$. From the latter branch the mixed polar-nematic state $\rho_{1,2}^\mathrm{pn}$ [green lines, Eq.~\eqref{pn}] emerges in pitchfork bifurcations at $c_0=0.091$ and $c_0=0.2425$. Solid (dashed) lines indicate stability (instability) w.r.t.\ uniform perturbations. Considering spatially extended systems, the state $\rho_{1,2}^\mathrm{n}$ first destabilises at a finite wave number in a wave instability (at $c_0=0.086$, marked by a square symbol and a blue ``W''). Note that for clarity the mirroring $\rho_{1,2}^\mathrm{pn}$-branch at negative $\rho_2$ is not included. Parameters are given in Eq.~\eqref{coeff}.
%
         % \ttuwe{If you like fig \& caption the main text still have to be adapted.}
          }
   \end{figure}

\subsection{Uniform states and their linear stability}
\label{sec:linstab}
First, we discussing the uniform steady states of Eqs.\,\eqref{eq6} that exist at the present parameters [Eqs.\,\eqref{coeff}] using the mean density $c_0$ as control parameter. The trivial state $\rho_1^\mathrm{t}=\rho_2^\mathrm{t}=0$ exists for all $c_0$. Two other branches of uniform steady states exist in certain ranges of $c_0$: The first branch bifurcates in a pitchfork bifurcation from the trivial state and exists for all $c_0>\beta_0/\beta_1=0.074$. It corresponds to purely nematic uniform states
\begin{eqnarray}
\label{nem}
\rho_1^\mathrm{n}=0,~~ \rho_2^\mathrm{n}=\pm\sqrt{(-\beta_0+c_0\beta_1)/\beta_3}.
\end{eqnarray}
 The second branch starts and ends in respective pitchfork bifurcations on the branch of nematic states  and exists for $c_0$ in the interval $0.091\leq c_0\leq 0.2425$. It represents a mixed polar-nematic uniform state with 
 \begin{eqnarray}
\label{pn}
\rho_2^\mathrm{pn}&=&(2\alpha_3)^{-1}[\alpha_2\pm \sqrt{\alpha_2^2-4\alpha_3(\alpha_0+c_0\alpha_1)}].\nonumber\\
\rho_1^\mathrm{pn}&=&\pm \sqrt{(\beta_2-\beta_4 \rho_2^\mathrm{pn})^{-1}[\beta_3(\rho_2^\mathrm{pn})^3+(\beta_0-c_0\beta_1)\rho_2^\mathrm{pn}]},
\end{eqnarray}
 Note that Eqs.~(\ref{nem}),\ref{pn}) are valid for dry systems as well as for wet systems, since the Marangoni flow induced by a uniform density distribution vanishes.
Fig.~\ref{F0} presents a bifurcation diagram featuring all uniform steady states. Both, polar order parameter $\rho_1$ and nematic order parameter $\rho_2$, are given as a function of $c_0$. The three-dimensional representation together with the projections of the bifurcation diagram onto the ($c_0, \rho_1$)- and ($c_0, \rho_2$)-planes well represent the connections between the different states and the symmetries of the system. In the case of a dry system, the bifurcation diagram in Fig.~\ref{F0} can be directly compared with a slice of the stability diagram obtained in Ref.~\cite{Denk20} for varying $c_0$ and $\alpha_2$. Indeed, at the value $\alpha_2=1.5$ here used the critical density for the nematic-polar transition is $c_0=0.091$ as obtained in Ref.~\cite{Denk20}.

%nematic order parameter $\rho_2^\mathrm{n}$ for the purely nematic state is shown in (a) with $c_0$ as a control parameter. The mixed polar-nematic state bifurcates from the purely nematic state at $c_0=0.091$ and merges with the latter at $c_0=0.2425$ as shown in Fig.\ref{F0}(b).

%The nematic order parameter $\rho_2^\mathrm{n}$ for the purely nematic state is shown in Fig.\ref{F0}(a) with $c_0$ as a control parameter. The mixed polar-nematic state bifurcates from the purely nematic state at $c_0=0.091$ and merges with the latter at $c_0=0.2425$ as shown in Fig.\ref{F0}(b).

%\ttuwe{I did not yet change the previous two sentences and the corresponding caption as I think a representation of $\rho_1$ for all states and of $\rho_2$ for all states in two respective panels will be much clearer. Then connect the panels by vertical lines that indicate important bifurcations as also done now. At the moment it is a curious mix. I would also include the family(s) of waves (then show averaged values). And use color and line style to distinguish curves. Even better would be a richly coloured 3d representation.} {\color{red} \tt a. Fig.\ref{F0} and caption have been amended.}
%\ttuwe{I still think that separating $\rho_{1,2}^\mathrm{n}$ and $\rho_{1,2}^\mathrm{pn}$ into two different panels makes no sense. Either one shows all $\rho_{1}$ in one panel and all $\rho_{2}$ in another or one choses a 3d-presentation. I had a go for the second version.}

To determine the linear stability of the uniform states that is already indicated in Fig.~\ref{F0}, we use the harmonic perturbation ansatz
 \begin{eqnarray}
 \label{ansatz}
   c&=&c_0+\varepsilon\tilde{c}e^{\lambda t+ikx},\nonumber\\
        \rho_1&=&\rho_1^\mathrm{s}+ \varepsilon\tilde{\rho}_1e^{\lambda t+ikx},\\
 \rho_2&=&\rho_2^\mathrm{s}+ \varepsilon\tilde{\rho}_2e^{\lambda t+ikx}, \nonumber
 \end{eqnarray}
 where the superscript s stands for any of the above introduced uniform steady states t, n, pn, and the quantities with tilde are the scaled perturbation amplitudes. Linearising Eqs.\,\eqref{eq6} in $\varepsilon\ll1$ about any of the uniform states one obtains an eigenvalue problem for the eigenvalues $\lambda$ and corresponding eigenvectors $\vecg{\rho}=(\tilde{c},\tilde{\rho}_1,\tilde{\rho}_2)^T$, namely,
\begin{eqnarray}
 \label{eig}
 \lambda \vecg{\rho}= (\tens{J}_0 + ik \tens{ J}_1 + k^2 \tens{J}_2) \vecg{\rho} = \tens{J} \vecg{\rho}.
 \end{eqnarray}
 Note that the real and imaginary part of $\lambda$ correspond to the growth rate and frequency of harmonic modes with wave number $k$, respectively. The propagation speed of travelling modes is given by $V=\text{Im}(\lambda)/k$.  The matrices within the Jacobian $\tens{J}$ are given by
 \begin{widetext}
 \begin{eqnarray}
 \label{JJ}
\tens{J}_0&=&
\left(
\begin{array}{ccc}
0 & 0 & 0\\
-\alpha_1\rho_1^\mathrm{s} &-(\alpha_0+c_0\alpha_1)+\alpha_2\rho_2^\mathrm{s}-\alpha_3(\rho_2^\mathrm{s})^2 & \alpha_2\rho_1^\mathrm{s}-2\alpha_3\rho_1^\mathrm{s}\rho_2^\mathrm{s} \\
\beta_1\rho_2^\mathrm{s} & 2\beta_2\rho_1^\mathrm{s}-2\beta_4\rho_2^\mathrm{s}\rho_1^\mathrm{s} & (-\beta_0+c_0\beta_1)-3\beta_3(\rho_2^\mathrm{s})^2-\beta_4(\rho_1^\mathrm{s})^2
\end{array}
\right),\nonumber\\
\tens{J}_1&=&
\left(
\begin{array}{ccc}
0 & -1 & 0\\
-\frac{1}{2} & 0& -\frac{1}{2}+\gamma_1\rho_2^\mathrm{s} \\
0& -\frac{1}{2}-\gamma_4\rho_2^\mathrm{s} & -(\gamma_4+\gamma_3)\rho_1^\mathrm{s}
\end{array}
\right),\qquad\tens{J}_2=
\left(
\begin{array}{ccc}
-(d+\text{Ma}\,c_0) & 0 & 0\\
-\text{Ma}\,\rho_1^\mathrm{s} & -d &0 \\
-\text{Ma}\,\rho_2^\mathrm{s} & 0 &  -(d+\gamma_2)
\end{array}
\right).
\end{eqnarray}
 \end{widetext}

   \begin{figure}[ht]
   \includegraphics[width=0.99\columnwidth]{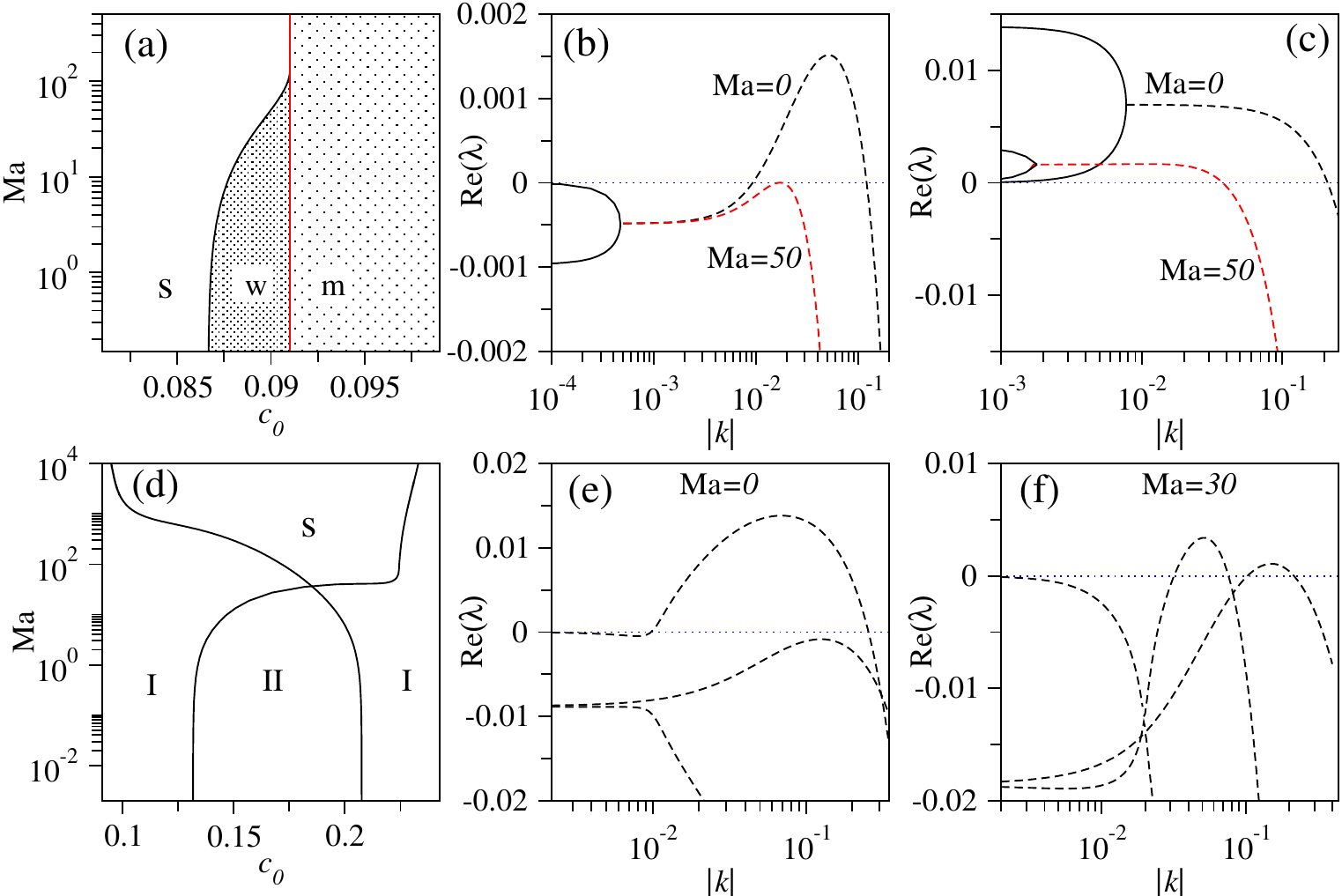}
   \caption{\label{F2} (a) Linear stability of the purely nematic uniform state $\rho_{1,2}^\mathrm{n}$ [Eq.\,\eqref{nem}] in the plane spanned by mean density $c_0$ and Marangoni parameter Ma.  Labels ``s'', ``w'' and ``m'' mark the linearly stable region, i.e., the region where all wave numbers are stable, the region of the wave instability (oscillatory instability of a finite band of wave numbers bound away from zero) and the region of mixed mode instability (unstable band of wave numbers $0\le k<k_c$ with real or complex eigenvalues depending on $k$), respectively, see main text.  Panels (b) and (c) give typical dispersion relations, i.e., growth rates $\text{Re}(\lambda)$  for the leading eigenvalues as a function of the wavenumber $k$, in region w at $c_0=0.09$ and in region m at $c_0=0.095$, respectively, at two different values of $\text{Ma}$ as indicated near each curve. Dashed (solid) lines correspond to complex (real) eigenvalues $\lambda$. 
     (d) Linear stability of the mixed polar-nematic uniform state $\rho_{1,2}^\mathrm{pn}$ [Eq.\,\eqref{pn}]. In the regions marked by $\text{I}$ [$\text{II}$] at most one [two] eigenvalues have a positive real part for any given $k$.   Panels (e) and (f) give dispersion relations in region $\text{I}$ at $\text{Ma}=0$, $c_0=0.12$ and region $\text{II}$ at $\text{Ma}=30$, $c_0=0.18$, respectively.}
   \end{figure}
  
 Note that the Jacobian matrix $\tens{J}$ is complex due to the term $ik \tens{J}_1$ that is related to self-propulsion. In consequence, for any wave number $k$, up to three different eigenvalues may exist, each either complex or real. More specifically, in the case of the trivial uniform state $\rho_{1,2}^{\mathrm{s}}=\rho_{1,2}^{\mathrm{t}}=0$ and in the case of the purely nematic state $\rho_{1,2}^{\mathrm{s}}=\rho_{1,2}^\mathrm{n}$ [Eq.~\eqref{nem}] the resulting characteristic polynomial $\det (\tens{J}-\lambda\mathbb{I})$ has only real coefficients. Therefore, all three eigenvalues are either real, or one eigenvalue is real and the remaining two form a complex conjugate pair. This feature is similar to isotropic systems where the Jacobian is always real.
 In contrast, for the mixed polar-nematic state with $\rho_{1,2}^{\mathrm{s}}=\rho_{1,2}^\mathrm{pn}$ the characteristic polynomial has complex coefficients resulting in three different, in general, complex eigenvalues. 

   The eigenvalue problem Eq.\,\eqref{eig} is solved numerically for various mean densities $c_0$ and Marangoni parameters $\text{Ma}$, to determine the regions of linear stability (where $\text{Re}(\lambda)<0$) for the three uniform steady states. Resulting stability diagrams for the nematic and the nematic-polar states in the plane $(c_0,\text{Ma})$ are given in Figs.~\ref{F2}~(a) and~\ref{F2}~(d), respectively, as explained in the following. As expected, the trivial state $\rho_1^\mathrm{t}=\rho_2^\mathrm{t}=0$ is (independently of Ma) linearly stable at small $c_0$ and becomes unstable at $c_0=\beta_0/\beta_1=0.074$. The instability is stationary and large-scale, i.e.\ the onset is at $k=0$.
At the instability threshold, the above-mentioned pitchfork bifurcation occurs where the uniform purely nematic state $\rho_{1,2}^\mathrm{n}$ [Eq.~\eqref{nem}] emerges. For slightly supercritical $c_0$, the leading eigenvalue of the trivial state is real and is given by $\lambda(k)=-\beta_0+c_0\beta_1-(d+\gamma_2)k^2-k^2/[4(\alpha_0+c_0(\alpha_1+\beta_1)-\beta_0)]+O(k^4)$.
   
Following the $\rho_{1,2}^\mathrm{n}$-branch for $\text{Ma}=0$ when gradually increasing $c_0$, we find a wave instability at $c_0=0.086$, i.e.\ a small-scale oscillatory instability, sometimes also called a finite-wavelength Hopf or oscillatory Turing instability. It is marked by a blue square symbol and the letter ``w'' in Fig.~\ref{F0}. Beyond this point the uniform nematic state is always unstable for sufficiently large domain sizes. At the wave bifurcation travelling nematic bands emerge as discussed in Ref.~\cite{Denk20}. Fig.~\ref{F2}~(b) shows the dispersion relation, represented by the real parts of the two leading eigenvalues in the parameter range between the onset of the wave instability and the pitchfork bifurcation where the uniform $\rho_{1,2}^\mathrm{pn}$-state emerges from $\rho_{1,2}^\mathrm{n}$. Dashed (solid) lines correspond to complex (real) eigenvalues $\lambda$. The dispersion relations are qualitatively similar in the entire region marked ``w'' in Fig.~\ref{F2}~(a). In this region, there exists a band of unstable wave numbers $k$, namely, the eigenvalue $\lambda$ is complex with a positive real part. The Marangoni flow suppresses the instability as shown in Fig.~\ref{F2}~(b).

At the onset of the wave instability (W-point in Fig.~\ref{F0}), the leading eigenvalue of the Jacobian is purely imaginary at some critical wave number $k_c$, i.e.\ $\lambda(k_c)=i\Omega$, where $V=\Omega/k_c$ is the speed of the propagating non-decaying wave that is at onset of infinitesimal amplitude. Further increasing $c_0$ beyond the W-point, the already unstable uniform nematic state at $c_0=0.091$ additionally undergoes the above described spatially uniform stationary instability where the $\rho_{1,2}^\mathrm{pn}$-branch emerges. Beyond this point, the dispersion relation for the $\rho_{1,2}^\mathrm{n}$-state features a real positive eigenvalue at small $k$ that connects to a pair of complex conjugate eigenvalues at larger $k$, see Fig.~\ref{F2}~(c). Together they form a continuous band of unstable wave numbers. 
Note that in the vicinity of $c_0=0.091$ the wavenumber band of the stationary instability becomes very narrow.  Exactly at $c_0=0.091$ it only contains the point $k=0$  implying that a this single parameter value the dispersion relation corresponds to a conserved-Hopf instability \cite{FrTh2023prl,GrTh2024c}.
% 
%{\color{red} \tt actually no, at $c_0=0.091$ the real part of the dispersion curve collapses to a point at $k=0$. For $c_0$ slightly above $0.091$ the real part of the dispersion curve re-appears with both values positive, as in Fig.\ref{F2}(c).}

We call the combination of real and complex eigenvalues above $c_0=0.091$ ``mixed mode'' and mark the corresponding region in Fig.~\ref{F2}~(a) by ``m''. Here, large-scale perturbations are monotonically unstable, while perturbations of  smaller scale are oscillatory unstable. However, both mode types form a common band of unstable wave numbers. Sometimes the transition point between real and complex modes within the band is called an ``exceptional point.'' Note however that such points only directly affect the primary bifurcations if they occur directly at zero crossing - then they correspond to a Takens-Bogdanov bifurcation. As they are detected in a linear analysis their potential influence on the nonlinear behaviour needs additional analysis. Inspecting Fig.~\ref{F2}~(a) we see that an inclusion of Marangoni flow suppresses the wave instability of the nematic state. It completely disappears at $\text{Ma}\approx 100$. Interestingly, this further implies that in contrast to the case studied in \cite{GrTh2024c}, here the conserved-Hopf instability appears as a codimension-2 instability exactly where the line separating the s- and w-region in Fig.~\ref{F2}~(a) ends.

It was shown already in Ref.~\cite{Denk20} for the dry case that the time evolution of the system in the mixed mode region is dominated by nematic-polar wave patterns characterised by spatially distributed polar and nematic order parameter fields. Analysing the linear stability of the uniform polar-nematic states $\rho_{1,2}^\mathrm{pn}$ [Eq.~\eqref{pn}] for a spatially extended system we find as expected that the state inherits the wave instability from the $\rho_{1,2}^\mathrm{n}$-branch it emerges from at $c_0=0.091$. A typical dispersion relation at moderate $c_0$ is given in Fig.~\ref{F2}~(e), and shows that the band of unstable wave numbers is bound away from zero. However, at lower and higher densities the unstable band can also start at $k=0$ (not shown).

The $\rho_{1,2}^\mathrm{pn}$-branch retains this instability till it ends at $c_0=0.2425$. It even acquires a second instability mode in an intermediate $c_0$-range, see the dispersion relation in Fig.~\ref{F2}~(f) shown for $c_0=0.18$ and $\text{Ma}=30$ close to the codimension-2 point in Fig.~\ref{F2}~(d), where the lines separating the regions ``I'' and ``II'' intersect. In these regions I and II there exist one and two wave number bands unstable wave modes, respectively.

From the stability diagram Fig.~\ref{F2}~(d) we see that an increase in the Marangoni parameter tends to suppress the wave instabilities: At $\text{Ma}\approx 100$ the II-region has been replaced by I-regions, i.e.\ the second wave mode has been suppressed, and a stable range of concentrations $c_0$ marked by ``s'' in Fig.~\ref{F2}~(d) has started to develop around $c_0=0.2$. 

 Note at the codimension-2 point in Fig.~\ref{F2}~(d), where the two lines meet that separate regions I and II, the two wave instabilities have a simultaneous onset. Adjusting parameters that control the corresponding critical wave numbers one might in the future be able to find resonances with interesting nonlinear behaviour.

\subsection{Travelling waves in the dry system}
\label{1d_dry}
%
%\ttuwe{I did not understand the discussion of ``may propagate through the purely nematic uniform state'' without corresponding profile pictures that prove it. Fig.4/5 only show density profiles, not $\rho_{1,2}$ profiles. Also the counting of waves is not clear to me, and in my opinion makes no sense: Either you talk about linear modes, then you have one in the upper row of Fig.~\ref{F2} and two in the lower row, or you talk about stationary nonlinear waves, then you need the full bifurcation diagram to talk about numbers. I tried to write something along the lines what you seem to have intended, pls check.}
%
%{\color{red} \tt Hmm... I am trying to understand your arguments. Ok, let me explain what I meant using the example of Fig.~\ref{F2}(e). Here we have one unstable band of the eignvectors $0.01<k<0.26$. Therefore, two branches of travelling waves exist, one befurcating at $k=0.01$ and another one at $k=0.26$.  I think this is exactly what you write later on, but this goes against this sentence where you say "The dispersion relations in (\color{red}{Figs.~\ref{F2}~(e) and~\ref{F2}~(f)  indicate that in regions $\text{I}$ and $\text{II}$ of Fig.~\ref{F2}~(d), respectively, one and two wave modes are unstable. Correspondingly, we expect that at least an equal number of branches of stationary travelling waves exists"} . So again,  in Fig.\ref{F2}~(e) only one mode is unstable - yes, but there exist two critical wave vectors, where two travelling wave branches bifurcate. }
%
The dispersion relations in Figs.~\ref{F2}~(e) and~\ref{F2}~(f)  indicate that in regions $\text{I}$ and $\text{II}$ of Fig.~\ref{F2}~(d), respectively, one and two wave modes are unstable. As both modes correspond to bands of unstable wave numbers bound away from zero, the one-mode [two-mode] case corresponds to two [four] bifurcations where branches of travelling waves emerge when using the domain size as control parameter. In situations as in the first row of Fig.~\ref{F2} there is one wave mode with two [Fig.~\ref{F2}~(b)] or one [Fig.~\ref{F2}~(c)]  bifurcations. Depending on the specific parameter values the bifurcations may be super- or subcritical, the branches may be connected or independent -- this is determined by the fully nonlinear behaviour. Directly at each bifurcation there exist harmonic waves of infinitesimal amplitude with propagation speed $V_c=\text{Im}(\lambda)/k_c$ and wavelength $L_c=2\pi/k_c$, where $k_c$ corresponds to the neutral mode, i.e.\ $\text{Re}(\lambda)(k_c)=0$. To gain knowledge about the branches of fully nonlinear travelling waves one has to resort to numerical methods.

Finite-amplitude travelling waves correspond to steady states in the co-moving frame that moves with the wave velocity $V$. Setting $c=c(x-Vt)$, $\rho_1=\rho_1(x-Vt)$ and $\rho_2=\rho_2(x-Vt)$ we rewrite Eqs.~\eqref{eq6} as a boundary value problem with periodic boundary conditions and an integral condition that controls the mean density $L^{-1}\int_0^{L} c(\chi)\,d\chi=c_0$, where $\chi=x-Vt$ is the phase of the wave in the co-moving frame. We then employ numerical path-continuation techniques \cite{KrauskopfOsingaGalan-Vioque2007,EGUW2019springer} bundled in the software package \textsc{auto07p} \cite{DoedelOldeman2009} to determine families of travelling waves.  As starting states we use the analytically known small-amplitude harmonic waves, i.e.\ Eq.~\eqref{ansatz} with eigenvalue, eigenvector and wavenumber of a neutral mode.

\begin{figure}[ht]
   \includegraphics[width=0.99\columnwidth]{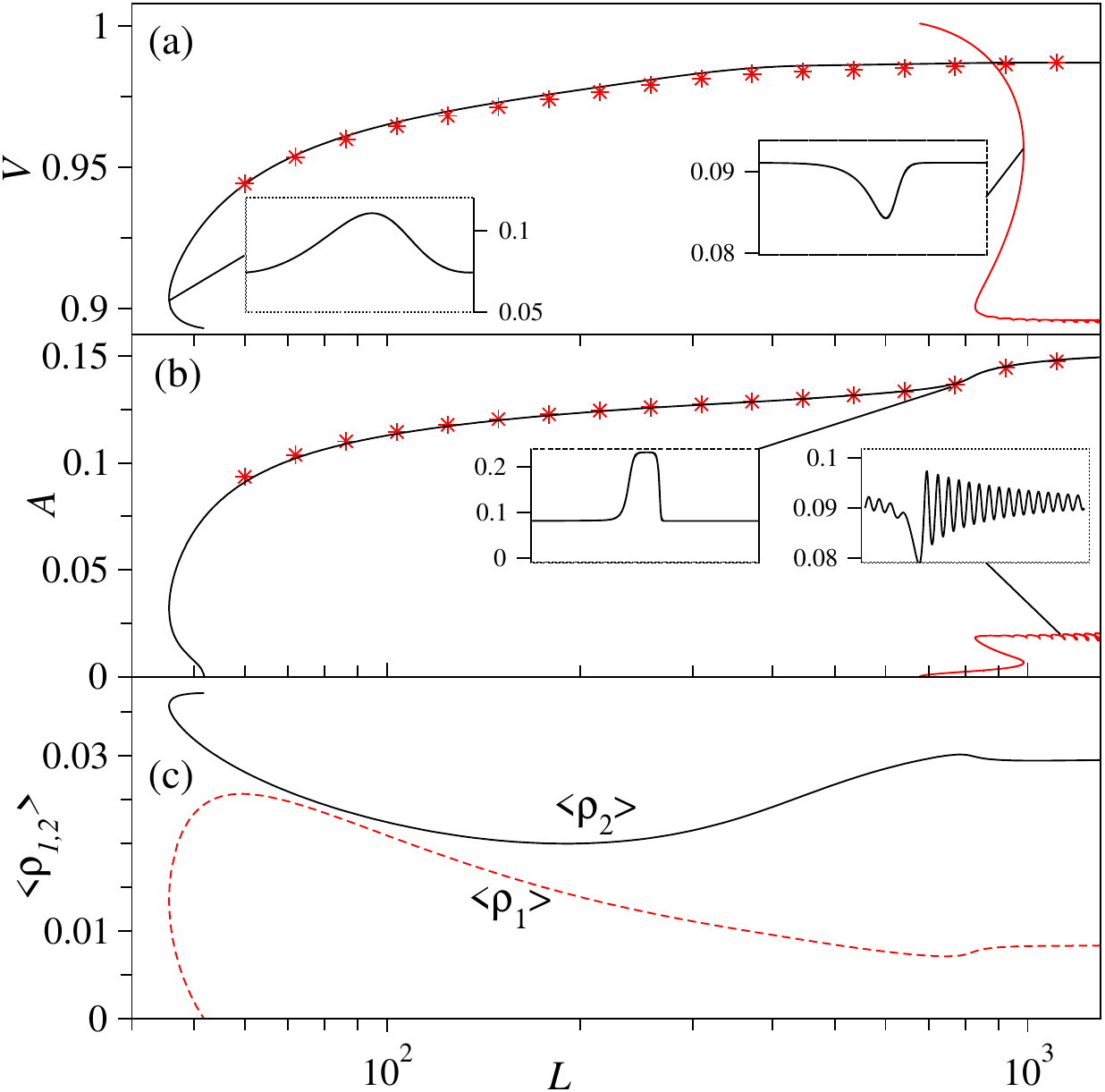}
   \caption{\label{F3} Two branches of travelling waves in a dry system ($\text{Ma}=0$) at fixed $c_0=0.09$ and $d=0.01$. Shown are (a) the wave speed $V$, (b) the amplitude of the density field $A=\text{max}(c)-\text{min}(c)$ as a function of the domain size $L$. Insets show the density profile at indicated points on the branches. Star symbols indicate stable wave states obtained by direct numerical time integration of Eqs.~\eqref{eq6}.  Panel~(c) gives the average polar and nematic fields $\langle \rho_{1,2}\rangle$ for the solution branch highlighted by the star symbols.}
   \end{figure}

First, we consider the dry system ($\text{Ma}=0$) at fixed mean density $c_0=0.09$ (where the polar-nematic state does not yet exist) and examine the wave bifurcation where a branch of travelling waves emerges from the uniform nematic state. As determined in section~\ref{sec:linstab}, for $c_0>0.086$ this uniform state is unstable in a certain range of wave numbers that depends on $c_0$. When we fix $c_0$ and use the domain size $L$ as control parameter, i.e.\ as principal continuation parameter, we encounter two disconnected branches of travelling waves that emerge where the dispersion relation in Fig.~\ref{F2}~(b) has its two zero crossings.
Fig.~\ref{F3} shows in panel~(a) the propagation speed $V$, in panel (b) the amplitude of the density field $A=\text{max}(c)-\text{min}(c)$, and in panel (c) the polar and mean nematic order parameter fields $\langle \rho_{1,2}\rangle = L^{-1}\int_0^L \rho_{1,2}(\chi),\,d\chi$ as a function of the domain size $L$. The wave with the larger [smaller] amplitude corresponds to a travelling density elevation [depression] both propagating in the uniform background, as shown in the insets of Fig.~\ref{F3}.
When numerically integrating Eqs.~(\ref{eq6}) in time using a naive continuation scheme as described in section~\ref{sec:2d}, at all $L$ the system converges to the state with the elevation (density hump, black line), as indicated by small star symbols in Fig.~\ref{F3}~(a) and (b). The travelling holes (red line) are never found in time simulations, which indicates that they are either unstable or linearly stable but with a very small basin of attraction.

One might expect that the two branches in Fig.~\ref{F3} will be connected in a parameter region close to onset of the underlying wave instability when the unstable wave number band is very small. This is, however, for the present system not the case. Instead a pre-existing branch of unstable nonlinear waves ``touches down'' to zero amplitude at the critical wavenumber at onset and roughly speaking splits into the two branches found above (not shown).
%  and that the found bifurcation diagram is the result of several re-organisations that occur with increasing $c_0$.

\begin{figure}[ht]
   \includegraphics[width=0.99\columnwidth]{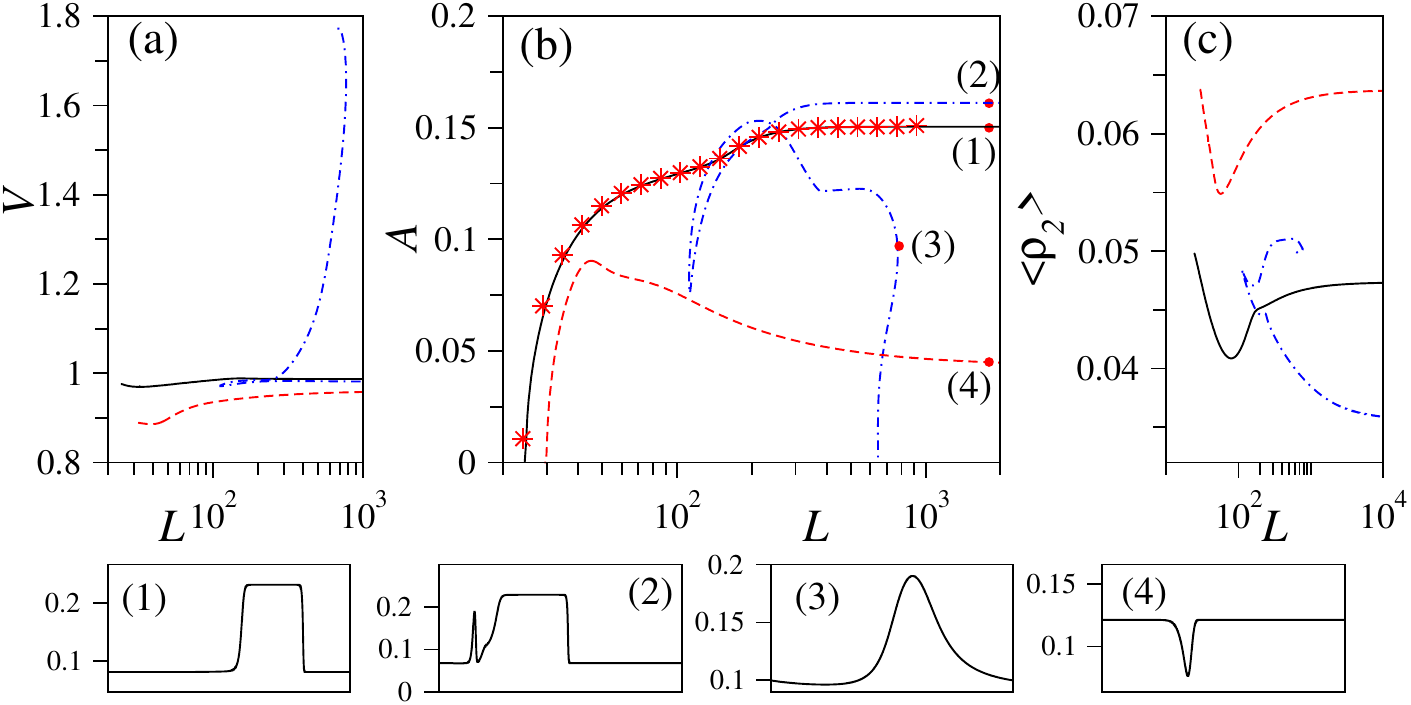}
   \caption{\label{F4} Three disconnected branches of finite-amplitude travelling waves for a dry system (Ma$=0$) at $c_0=0.12$ and $d=0.01$. (a) Wave speed $V$, (b) amplitude of the density field $A=\text{max}(c)-\text{min}(c)$ and (c) average nematic order parameter field $\langle \rho_2\rangle$ as a function of domain size $L$. The four red filled circles marked (1) to (4) in panel (b) indicate loci of density profiles shown in the bottom panels. Star symbols indicate stable wave states obtained by direct numerical time integration of Eqs.~\eqref{eq6}.}
   \end{figure}

   Second, we look at the dry system at the larger mean density $c_0=0.12$ where also the uniform polar-nematic state exists. The linear stability results then indicate that overall three types of small-amplitude travelling wave states may emerge from the uniform states (one from purely nematic and two from mixed polar-nematic). The corresponding zero crossings of the dispersion relations (Fig.~\ref{F2}) correspond to three wave bifurcations and may give rise to up to three disconnected branches of finite-amplitude waves (the exact number depends on the fully nonlinear behaviour). At $c_0=0.12$ we find three disconnected branches, see in Figs.~\ref{F4}~(a)-(c). The first one bifurcates at the critical $k$ from the purely nematic state (red dashed line) and the other two at the critical $k$'s for the mixed polar-nematic state (solid black and dot-dashed blue line). The dot-dashed branch undergoes two saddle-node bifurcations, however, at large domain sizes three clearly distinguished travelling waves remain. They correspond to a single-hump wave (solid black line) as shown in Fig.~\ref{F4}(1,3), a double-hump wave (blue dot-dashed line) as shown in Fig.~\ref{F4}(2), and a travelling-hole state (red dashed line) as shown in Fig.~\ref{F4}(4). The double-hump state corresponds to a large travelling density cluster followed at a small distance by a small satellite peak. Direct numerical time integration of Eqs.~\eqref{eq6} using the naive continuation algorithm as described above only converges to single-hump waves (as indicated by star symbols in \ref{F4}~(a)-(c)) indicating their linear stability. The other types of travelling waves were not found in the time simulations.
   
For the purpose of simplicity, we do not discuss the more complicated bifurcation diagram for travelling waves in region $\text{II}$ of Fig.~\ref{F2}~(d), where five wave bifurcations occur when using $L$ as control parameter: four branches emerge from the mixed polar-nematic state and one from the purely nematic state.

\subsection{Effect of the Marangoni flow on the travelling waves }
\label{1d_wet}
\begin{figure}[ht]
   \includegraphics[width=0.99\columnwidth]{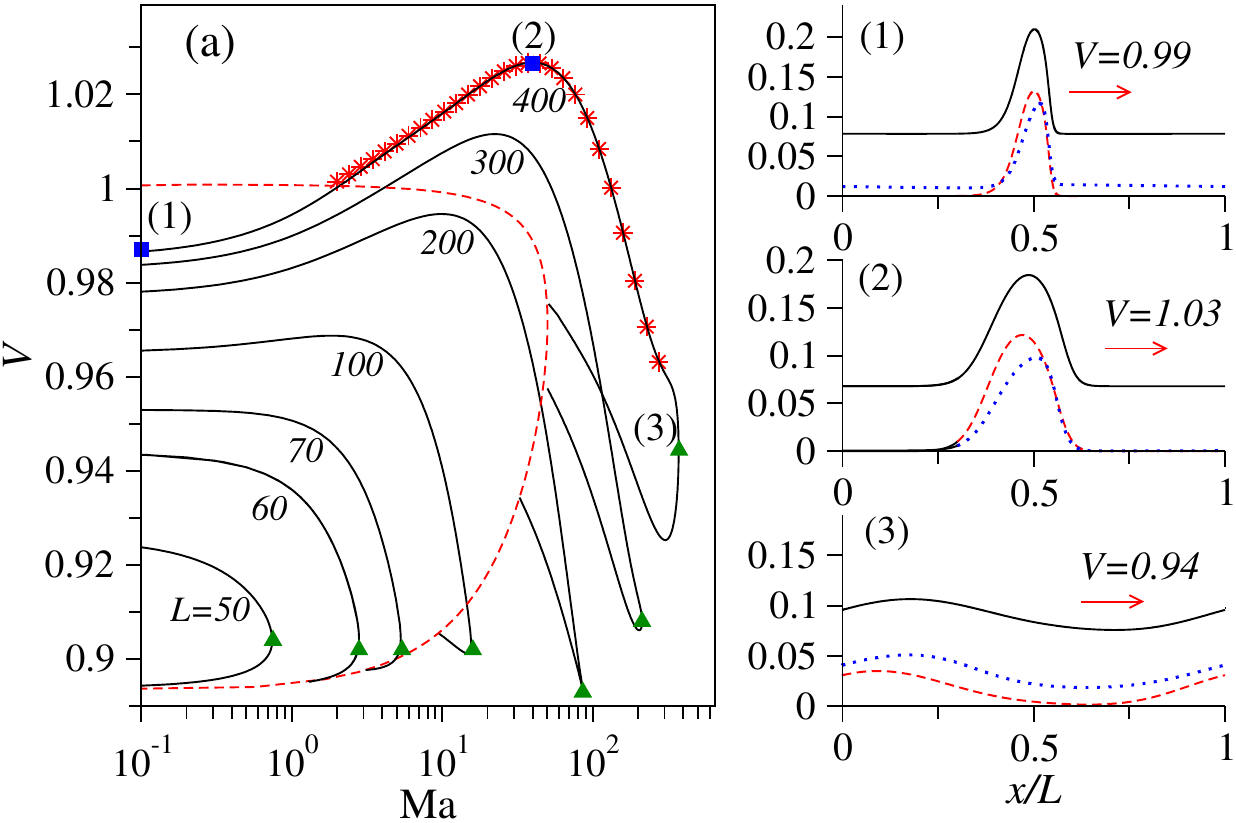}
   \caption{\label{F5}  (a) Speed of nonlinear travelling waves as a function of $\text{Ma}$ for $c_0=0.09$ at various fixed domain sizes $L$ as indicated at the curves. Star symbols indicate linearly stable solutions found using direct numerical integration of Eqs.~\eqref{eq6} for $L=400$. Square symbols with labels (1), (2) and (3) on the same branch indicate representative states that are shown in the panels on the right where black solid, red dashed and blue dotted lines show $c$, $\rho_1$ and $\rho_2$ profiles, respectively. In (a) saddle-node bifurcation points are marked by green triangles, and the dashed red line corresponds to the loci of wave bifurcations in the (Ma,$V$)-plane where branches of travelling waves emerge from the purely nematic state [Eq.~\eqref{nem}].}
   \end{figure}

Finally, we transition in a controlled manner from the dry to the wet system and analyse the effect of an increasing Marangoni flow on the propagation of polar-nematic waves. 
%\ttuwe{we still need to clearly establish that the waves are ``polar-nematic waves'' by showing $c_0,\rho_1,\rho_2$ profiles of the stable wave states - at the moment it is just a claim.} 
%
To do so, we follow the linearly stable nonlinear travelling waves found in section~\ref{1d_dry} when increasing the Marangoni parameter $\text{Ma}$ from zero at various fixed domain sizes $L$. Fig.~\ref{F5}~(a) shows resulting branches of travelling wave states for five values $L=50,60,70,100,200,300,400$ at fixed $c_0=0.09$ (as in Fig.~\ref{F3}).
 
  \begin{figure}[ht]
   \includegraphics[width=0.9\columnwidth]{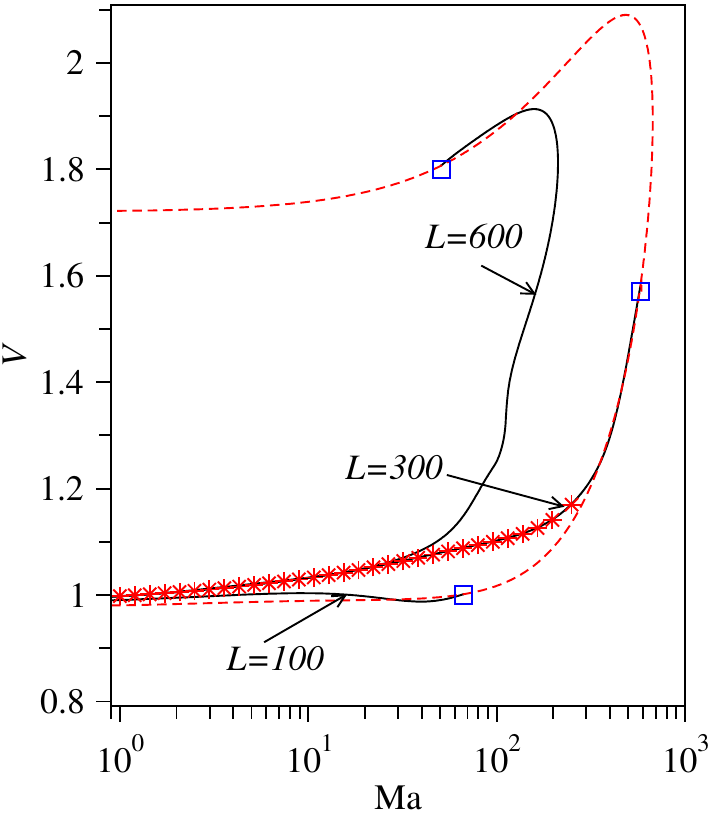}
   \caption{\label{F6}  Speed of travelling waves of different wavelength $L$ as a function of $\text{Ma}$ for $c_0=0.12$. Star symbols indicate stable wave states found using direct numerical integration of Eqs.\,\eqref{eq6} in a periodic domain of size $L=300$. The dashed red line corresponds to the loci of wave bifurcations in the (Ma,$V$)-plane where branches of travelling waves emerge from the mixed polar-nematic uniform state [Eq.~\eqref{pn}]. For the shown $L$ these points are indicated by empty square symbols.}
 \end{figure}

Interestingly, the speed of the waves of larger periods, $L>100$, depends nonmonotonically on $\text{Ma}$. In complete agreement with the 2D case discussed at Fig.~\ref{F1}, the speed of the waves on the stable part of the branches increases with increasing $\text{Ma}$, reaches a maximal value at a certain value $\text{Ma}_\mathrm{max}(L)$, and then decreases again until the branch reaches the saddle-node bifurcation, marked by a green triangle. There, the stable part of the branch annihilates with the unstable part.

A typical scenario is for $L=400$ further illustrated by the density, polarisation, and nematic profiles on the right of Fig.~\ref{F5}: a stable wave in a dry system [state (1)] propagates at a speed of $V=0.99$, the speed slightly increases to $V=1.03$ at $\text{Ma}=42$ [state (2)], and then decreases to $V=0.95$ at the saddle-node bifurcation at $\text{Ma}=370$ [state (3)]. The solitary waves (1) and (2) show pronounced peaks of density as well as of polar and nematic order, and move in a nearly isotropic background, i.e.\ without polar or nematic order.

Note that the 2D travelling waves of maximal speed shown in Fig.~\ref{F1}~(c) have an effective wavelength of about $400$, which corresponds to the branch with  $L=400$ in Fig.~\ref{F5}~(a). Stable waves are not found beyond the saddle-node bifurcation, as confirmed by direct numerical simulations of Eqs.~\eqref{eq6}, indicated by star symbols in Fig.~\ref{F5}~(a). At the saddle-node bifurcation, the branch of travelling wave states becomes unstable and turns toward smaller $\text{Ma}$. Finally, it terminates in a subcritical wave bifurcation on the dashed red line that indicates the loci of wave bifurcations in the (Ma,$V$)-plane. There, a continuous spectrum of branches of travelling waves of different $L$ emerges from the purely nematic state [Eq.~\eqref{nem}]. It is obtained by continuing the zero crossing $\text{Re}(\lambda)=0$ of Eqs.~\eqref{eig} in parameter space when gradually changing $L=2\pi/k$.

The speed $V$ of the stable waves with the relatively small wavelength $L=50$ monotonically decreases with increasing $\text{Ma}$ until the saddle-node bifurcation point is reached. If the Marangoni parameter is increased beyond the saddle-node point, in a time simulation the finite-amplitude wave relaxes to a uniform solution. 

Interestingly, at larger mean densities, e.g.\ $c_0=0.12$ and moderately small wavelength $L\lesssim300$, we find that the destruction of the wave by Marangoni flow may follow a different scenario. As shown in Fig.~\ref{F6}, e.g.\ for $L=300$, the speed of stable waves can also monotonically increase with increasing $\text{Ma}$ and by a much larger percentage as at lower $c_0$), until the point where the branch of the wave states ends in a wave bifurcation on the mixed polar-nematic uniform state.
In this scenario, the wave amplitude gradually decreases until it eventually reaches zero at the supercritical wave bifurcation on the dashed red line.  This is in contrast to the scenario of Fig.~\ref{F5}, where the finite-amplitude waves bifurcate from the trivial purely nematic state (given by the dashed line) via a sub-critical wave bifurcation. 
%\ttuwe{trivial or nematic state?}

%\ttuwe{I think one should add a discussion of what happens in large domains, by adding some really large-scale simulations, ie periods of several 1000. Will certain wavelength be selected, will the dynamics be chaotic, etc. Such questions typically arise when showing bif.\ diagrams.}

%{\tt andrey: I did a few simulations with a 1D system in a large domain of size $2000$ for $c_0=0.12$. I did not find any chaos with or without Marangoni flow. Movies are attached: Ma0.mp4 dynamics at $Ma=0$ with random initial conditions without bias. Ma0-bias.mp4 - same as Ma0.mp4, but with a small bias in the initial conditions for the field $\rho_1$ in the positive direction. Ma100-bias.mp4 - same as  Ma0-bias.mp4 but with $Ma=100$.}

      % \ttuwe{up to here}

   \section{Conclusion}
   \label{conc}
   We have studied the effect of self-induced Marangoni flow on the propagation of periodic and solitary density waves in a system of self-propelled agents with aligning interactions that move on the planar surface of a viscous liquid film. Thereby, the agents act as an insoluble surfactant in the purely entropic regime, i.e.\ with increasing density the surface tension of the liquid decreases linearly. This implies that emerging concentration gradients result in Marangoni forces that cause self-induced Marangoni flows.  These influence the dynamics of the agents in such a way that advective flows in the liquid film and the dynamics of the active agents are intricately coupled. In this way we have been able to quantitatively compare the dry case (no self-induced liquid flow) and the wet case  (self-induced liquid flow). 

The dynamics of a dry system with local nematic alignment is described by a truncated Boltzmann equation derived in Ref.~\cite{Denk20,Peshkov12}. Here we have adapted the model for wet systems by incorporating an advection term that results from the Marangoni flow. After establishing a simplified dynamical model that couples density, polar order parameter and nematic order parameter fields, we have focused of the case of a long-wave approximation where the local flow field is proportional to the local density gradient. To characterise the degree of departure from the dry regime, we have introduced a Marangoni parameter that measures the relative strength of the Marangoni flow as compared to the self-propulsion speed of the agents. 

We have then chosen agent interaction parameters that correspond to a regime where stable travelling density waves of mixed polar-nematic order are known to exist in the dry limiting case. As the Marangoni parameter is gradually increased, these waves broaden in the direction of propagation, while their speed either decreases or increases monotonically or changes nonmonotonically depending on the mean concentration and wavelength. In addition, in the 2D case the density distribution homogenises in the direction transverse to the wave propagation, leading to the formation of almost perfectly parallel travelling density bands. We emphasise that such bands are well known to form in the absence of the Marangoni flow in dry as well as wet active-matter systems   \cite{Vicsek1995,ToTu1998pre,ToTR2005ap,SoTa2013prl,SoCT2015prl,Lushi2018}. As the strength of the Marangoni flow is further increased, the waves are destroyed and the system relaxes to a uniform polar or mixed polar-nematic state depending on the mean density.  Depending on the wavelength of the nonlinear waves this occurs either discontinuously via a saddle-node bifurcation of finite-amplitude travelling waves or continuously via a gradual decrease of the wave amplitude till it reaches zero at a wave bifurcation of a uniform state. The two scenarios correspond to cases of subcritical and supercritical wave bifurcations, respectively, that result in different nonequilibrium phase behaviour.

%{\tt andrey: I went through the papers by Tonner, Tailleur and others on the continuous flocking models and did not find any reference to the type of bifurcations to travelling waves or any bifurcation analysis. It looks like the emphasis of the earlier papers was on the statistical properties of phase transitions in the active-matter systems. The main conclusion was that even in two dimensions, the proper phase transition to flocking is possible, unlike in the equilibrium thermodynamic, where no phase transitions are possible in 2D systems with local interactions (Ising model). Basically, I failed to find any study on the effect of the Marangoni flow on the travelling waves in active matter systems. Probably the only paper, which is sort of comes close to what we do here is the \cite{Lushi2018}, where an autochemotactic system of self-propelled agents was studied. The main conclusion of this paper is: "quaisi-one-dimensional density bands appear in the system due to chemotactic instability. These bands transition to travelling pulses  under an external chemotactic gradient. "}

Our results have shown strong parallels between wet active-matter systems with local alignment and reaction-diffusion systems with advection \cite{Rongy2006,Rongy2007}. In both types of systems the propagation of fronts and waves is similarly influenced by the addition of hydrodynamic flow, i.e.\ by ``making the system wet.'' Specifically, in autocatalytic chemical reaction systems, the Marangoni flow causes wave fronts to widen and increases their propagation speed. Similarly, in active-matter systems composed of self-propelled particles, we have observed that the propagation speed of polar-nematic waves increases with the Marangoni parameter, provided the wavelength is sufficiently large. The widening of the waves we have observed can also be attributed to the Marangoni advection flow, which moves away from regions of higher concentration. This analogy between the two systems can be understood from a phenomenological perspective: in the limit of small density variations, the dynamics of active-matter systems can be described by a closed set of reaction-advection-diffusion equations governing the dominant order parameter fields.

\section*{Author contribution statement}
All authors contributed equally to conceptualisation, writing and literature research. (AP) numerical simulations, bifurcation analysis and figure preparation. (UT) figure preparation and analysis.

\bmhead{Acknowledgements}

UT would like to thank the Kavli Institute for Theoretical Physics (KITP), Santa Barbara, for support and hospitality during the programme \textit{Active Solids} where part of the work was undertaken. This research was partially supported by the Australian Government through the Australian Research Council's Discovery Projects funding scheme (project DP250100868) and by grant NSF PHY-2309135 to KITP. 

\bmhead{Data availability}
All data generated or analysed during this study are included in this published article [and its supplementary information files]. Source codes are available from AP on request.

\appendix
\section{Closed system of hydrodynamic equations for the leading order parameter fields in the presence of advective flow}
\label{appendix}
Consider the two-dimensional motion of self-propelled agents in the presence of an advection flow field $\vec{U}(c)$ which only depends on the local concentration of the agents $c$ and is insensitive to their local orientation. Following \cite{Bertin2009,Peshkov12,Denk20} we write the Boltzmann equation~(\ref{eq1}) in the angular Fourier space in terms of the order parameter fields $\rho_q$ defined by equation~(\ref{eqorderParameters})
\begin{eqnarray}
\label{app_eq1}
\partial_t \rho_q+\vec{\nabla}\cdot (\vec{U}\rho_q) +\frac{v_0}{2} (\partial \rho_{q-1} +\partial^* \rho_{q+1})=d\partial \partial^* \rho + R_q \rho_q+\sum_{n=-\infty}^\infty I_{n,q}\rho_n\rho_{q-n}.
\end{eqnarray}
Here, $\partial =\partial_x +i\partial_y$, $R_q<0$ describes random re-orientation due to rotational diffusion and $I_{n,q}$ is the collision matrix in the angular Fourier space. Note that $R_q$ and $I_{n,q}$ depend on microscopic parameters such as the interaction distance, the alignment rule  during two-particle collisions and the statistics of rotational diffusion.
For example, as shown in \cite{Denk20}, $R_q=-\lambda (1-e^{-(q\sigma^2/2)})$, where $\lambda$ is the rate at which the agents randomly change their orientation from $\theta$ to $\theta +\eta$, where $\eta$ is normally distributed with zero mean and standard deviation $\sigma$. The collisions matrix $I_{n,q}$ consists of two parts: the first part describes polar alignment between two agents that collide at an acute angle and the second part corresponds to antipolar alignment when the collision occurs at an obtuse angle. For simplicity of discussion, we do not show the expression for $I_{n,q}$ in terms of the specific rules for collision statistics.

Assuming $\rho_{q>4}=0$, we obtain for $\rho_{3,4}$ the kinetic equations
\begin{align}
  \label{app_eq2}
  \begin{split}
\partial_t \rho_{3}+\vec{\nabla}\cdot (\vec{U}\rho_3) +\frac{v_0}{2} (\partial \rho_2 +\partial^* \rho_4)&=d\vec{\nabla}^2 \rho_3 + R_3 \rho_3 +  \sum_{n=-1}^4I_{n,3}\rho_n \rho_{3-n},\\
\partial_t \rho_{4}+\vec{\nabla}\cdot (\vec{U}\rho_4) +\frac{v_0}{2} (\partial \rho_3 )&=d\vec{\nabla}^2 \rho_4 + R_4 \rho_4 +  \sum_{n=0}^4I_{n,4}\rho_n \rho_{4-n}.
\end{split}
\end{align}
After applying the scaling~(\ref{scale}) to equations~(\ref{app_eq2}) we retain only terms up to the order $\epsilon^3$. At this order, the concentration $c$ is replaced with its mean value $c_0$. The relationship between $\rho_{4}$ and $(c, \rho_{1,2})$ is strictly local, while $\rho_3$ also depends on the gradient of $\rho_2$, namely,
\begin{equation}
\label{app_eq3}
\rho_4=-\frac{I_{2,4}\rho_2^2}{R_4+c_0(I_{0,4}+I_{4,4})},~~\rho_3= \frac{(v_0/2) \partial \rho_2-I_{1,3}\rho_1\rho_2}{R_3+c_0(I_{0,3}+I_{3,3})}.
\end{equation}
Note that the closure relations~(\ref{app_eq3}) do not depend on the advecting surface flow $\vec{U}$, and therefore exactly coincide with those derived in Refs.~\cite{Peshkov12,Denk20}. In consequence, the resulting closed system of hydrodynamic equations for $c$, $\rho_{1,2}$ for a wet system correspond to the hydrodynamic equations derived in Refs.~\cite{Peshkov12,Denk20} each extended by the advection term $\vec{\nabla}\cdot (\vec{U}\rho_q)$ with $(q=0,1,2)$ corresponding to the self-induced Marangoni flow.

%\bibliography{bibfile}

%\bibliography{bibfile}% common bib file

\begin{thebibliography}{46}
\ifx \bisbn   \undefined \def \bisbn  #1{ISBN #1}\fi
\ifx \binits  \undefined \def \binits#1{#1}\fi
\ifx \bauthor  \undefined \def \bauthor#1{#1}\fi
\ifx \batitle  \undefined \def \batitle#1{#1}\fi
\ifx \bjtitle  \undefined \def \bjtitle#1{#1}\fi
\ifx \bvolume  \undefined \def \bvolume#1{\textbf{#1}}\fi
\ifx \byear  \undefined \def \byear#1{#1}\fi
\ifx \bissue  \undefined \def \bissue#1{#1}\fi
\ifx \bfpage  \undefined \def \bfpage#1{#1}\fi
\ifx \blpage  \undefined \def \blpage #1{#1}\fi
\ifx \burl  \undefined \def \burl#1{\textsf{#1}}\fi
\ifx \doiurl  \undefined \def \doiurl#1{\url{https://doi.org/#1}}\fi
\ifx \betal  \undefined \def \betal{\textit{et al.}}\fi
\ifx \binstitute  \undefined \def \binstitute#1{#1}\fi
\ifx \binstitutionaled  \undefined \def \binstitutionaled#1{#1}\fi
\ifx \bctitle  \undefined \def \bctitle#1{#1}\fi
\ifx \beditor  \undefined \def \beditor#1{#1}\fi
\ifx \bpublisher  \undefined \def \bpublisher#1{#1}\fi
\ifx \bbtitle  \undefined \def \bbtitle#1{#1}\fi
\ifx \bedition  \undefined \def \bedition#1{#1}\fi
\ifx \bseriesno  \undefined \def \bseriesno#1{#1}\fi
\ifx \blocation  \undefined \def \blocation#1{#1}\fi
\ifx \bsertitle  \undefined \def \bsertitle#1{#1}\fi
\ifx \bsnm \undefined \def \bsnm#1{#1}\fi
\ifx \bsuffix \undefined \def \bsuffix#1{#1}\fi
\ifx \bparticle \undefined \def \bparticle#1{#1}\fi
\ifx \barticle \undefined \def \barticle#1{#1}\fi
\bibcommenthead
\ifx \bconfdate \undefined \def \bconfdate #1{#1}\fi
\ifx \botherref \undefined \def \botherref #1{#1}\fi
\ifx \url \undefined \def \url#1{\textsf{#1}}\fi
\ifx \bchapter \undefined \def \bchapter#1{#1}\fi
\ifx \bbook \undefined \def \bbook#1{#1}\fi
\ifx \bcomment \undefined \def \bcomment#1{#1}\fi
\ifx \oauthor \undefined \def \oauthor#1{#1}\fi
\ifx \citeauthoryear \undefined \def \citeauthoryear#1{#1}\fi
\ifx \endbibitem  \undefined \def \endbibitem {}\fi
\ifx \bconflocation  \undefined \def \bconflocation#1{#1}\fi
\ifx \arxivurl  \undefined \def \arxivurl#1{\textsf{#1}}\fi
\csname PreBibitemsHook\endcsname

%%% 1
\bibitem[\protect\citeauthoryear{Ramaswamy}{2010}]{Ramaswamy2010}
\begin{barticle}
\bauthor{\bsnm{Ramaswamy}, \binits{S.}}:
\batitle{The mechanics and statistics of active matter}.
\bjtitle{Annu. Rev. Condens. Matter Phys.}
\bvolume{1},
\bfpage{323}--\blpage{345}
(\byear{2010})
\end{barticle}
\endbibitem

%%% 2
\bibitem[\protect\citeauthoryear{Marchetti et~al.}{2013}]{Marchetti2013}
\begin{barticle}
\bauthor{\bsnm{Marchetti}, \binits{M.C.}},
\bauthor{\bsnm{Joanny}, \binits{J.F.}},
\bauthor{\bsnm{Ramaswamy}, \binits{S.}},
\bauthor{\bsnm{Liverpool}, \binits{T.B.}},
\bauthor{\bsnm{Prost}, \binits{J.}},
\bauthor{\bsnm{Rao}, \binits{M.}},
\bauthor{\bsnm{Simha}, \binits{R.A.}}:
\batitle{Hydrodynamics of soft active matter}.
\bjtitle{Rev. Mod. Phys.}
\bvolume{85},
\bfpage{1143}--\blpage{1189}
(\byear{2013})
\end{barticle}
\endbibitem

%%% 3
\bibitem[\protect\citeauthoryear{B{\"a}r et~al.}{2020}]{Baer2020}
\begin{barticle}
\bauthor{\bsnm{B{\"a}r}, \binits{M.}},
\bauthor{\bsnm{Grossmann}, \binits{R.}},
\bauthor{\bsnm{Heidenreich}, \binits{S.}},
\bauthor{\bsnm{Peruani}, \binits{F.}}:
\batitle{Self-propelled rods: Insights and perspectives for active matter}.
\bjtitle{Annu. Rev. Condens. Matter Phys.}
\bvolume{11},
\bfpage{441}--\blpage{466}
(\byear{2020})
\end{barticle}
\endbibitem

%%% 4
\bibitem[\protect\citeauthoryear{Vicsek et~al.}{1995}]{Vicsek1995}
\begin{barticle}
\bauthor{\bsnm{Vicsek}, \binits{T.}},
\bauthor{\bsnm{Czir\'ok}, \binits{A.}},
\bauthor{\bsnm{Ben-Jacob}, \binits{E.}},
\bauthor{\bsnm{Cohen}, \binits{I.}},
\bauthor{\bsnm{Shochet}, \binits{O.}}:
\batitle{Novel type of phase transition in a system of self-driven particles}.
\bjtitle{Phys. Rev. Lett.}
\bvolume{75},
\bfpage{1226}--\blpage{1229}
(\byear{1995})
\end{barticle}
\endbibitem

%%% 5
\bibitem[\protect\citeauthoryear{Toner and Tu}{1998}]{ToTu1998pre}
\begin{barticle}
\bauthor{\bsnm{Toner}, \binits{J.}},
\bauthor{\bsnm{Tu}, \binits{Y.}}:
\batitle{Flocks, herds, and schools: a quantitative theory of flocking}.
\bjtitle{Phys. Rev. E}
\bvolume{58},
\bfpage{4828}--\blpage{4858}
(\byear{1998})
\doiurl{10.1103/PhysRevE.58.4828}
\end{barticle}
\endbibitem

%%% 6
\bibitem[\protect\citeauthoryear{Toner et~al.}{2005}]{ToTR2005ap}
\begin{barticle}
\bauthor{\bsnm{Toner}, \binits{J.}},
\bauthor{\bsnm{Tu}, \binits{Y.}},
\bauthor{\bsnm{Ramaswamy}, \binits{S.}}:
\batitle{Hydrodynamics and phases of flocks}.
\bjtitle{Ann. Phys.}
\bvolume{318},
\bfpage{170}--\blpage{244}
(\byear{2005})
\doiurl{10.1016/j.aop.2005.04.011}
\end{barticle}
\endbibitem

%%% 7
\bibitem[\protect\citeauthoryear{Solon and Tailleur}{2013}]{SoTa2013prl}
\begin{barticle}
\bauthor{\bsnm{Solon}, \binits{A.P.}},
\bauthor{\bsnm{Tailleur}, \binits{J.}}:
\batitle{Revisiting the flocking transition using active spins}.
\bjtitle{Phys. Rev. Lett.}
\bvolume{111},
\bfpage{078101}
(\byear{2013})
\doiurl{10.1103/physrevlett.111.078101}
\end{barticle}
\endbibitem

%%% 8
\bibitem[\protect\citeauthoryear{Solon et~al.}{2015}]{SoCT2015prl}
\begin{barticle}
\bauthor{\bsnm{Solon}, \binits{A.P.}},
\bauthor{\bsnm{Chat{\'e}}, \binits{H.}},
\bauthor{\bsnm{Tailleur}, \binits{J.}}:
\batitle{From phase to microphase separation in flocking models: {T}he
  essential role of nonequilibrium fluctuations}.
\bjtitle{Phys. Rev. Lett.}
\bvolume{114},
\bfpage{068101}
(\byear{2015})
\doiurl{10.1103/PhysRevLett.114.068101}
\end{barticle}
\endbibitem

%%% 9
\bibitem[\protect\citeauthoryear{Couzin et~al.}{2005}]{Couzin2005}
\begin{barticle}
\bauthor{\bsnm{Couzin}, \binits{I.D.}},
\bauthor{\bsnm{Krause}, \binits{J.}},
\bauthor{\bsnm{Franks}, \binits{N.R.}},
\bauthor{\bsnm{Levin}, \binits{S.A.}}:
\batitle{Effective leadership and decision-making in animal groups on the
  move}.
\bjtitle{Nature}
\bvolume{433},
\bfpage{513}--\blpage{516}
(\byear{2005})
\end{barticle}
\endbibitem

%%% 10
\bibitem[\protect\citeauthoryear{Lushi et~al.}{2018}]{Lushi2018}
\begin{barticle}
\bauthor{\bsnm{Lushi}, \binits{E.}},
\bauthor{\bsnm{Goldstein}, \binits{R.E.}},
\bauthor{\bsnm{Shelley}, \binits{M.J.}}:
\batitle{Nonlinear concentration patterns and bands in autochemotactic
  suspensions}.
\bjtitle{Phys. Rev. E}
\bvolume{98},
\bfpage{052411}
(\byear{2018})
\doiurl{10.1103/PhysRevE.98.052411}
\end{barticle}
\endbibitem

%%% 11
\bibitem[\protect\citeauthoryear{Schaller et~al.}{2010}]{Schaller2010}
\begin{barticle}
\bauthor{\bsnm{Schaller}, \binits{V.}},
\bauthor{\bsnm{Weber}, \binits{C.}},
\bauthor{\bsnm{Semmrich}, \binits{C.}},
\bauthor{\bsnm{Frey}, \binits{E.}},
\bauthor{\bsnm{Bausch}, \binits{A.R.}}:
\batitle{Polar patterns of driven filaments}.
\bjtitle{Nature}
\bvolume{467},
\bfpage{73}--\blpage{77}
(\byear{2010})
\end{barticle}
\endbibitem

%%% 12
\bibitem[\protect\citeauthoryear{Bertin et~al.}{2009}]{Bertin2009}
\begin{barticle}
\bauthor{\bsnm{Bertin}, \binits{E.}},
\bauthor{\bsnm{Droz}, \binits{M.}},
\bauthor{\bsnm{Gr\'egoire}, \binits{G.}}:
\batitle{Hydrodynamic equations for self-propelled particles: microscopic
  derivation and stability analysis}.
\bjtitle{J. Phys. A: Math. Theor.}
\bvolume{42},
\bfpage{445001}
(\byear{2009})
\end{barticle}
\endbibitem

%%% 13
\bibitem[\protect\citeauthoryear{Peshkov et~al.}{2012}]{Peshkov12}
\begin{barticle}
\bauthor{\bsnm{Peshkov}, \binits{A.}},
\bauthor{\bsnm{Aranson}, \binits{I.S.}},
\bauthor{\bsnm{Bertin}, \binits{E.}},
\bauthor{\bsnm{Chat\'e}, \binits{H.}},
\bauthor{\bsnm{Ginelli}, \binits{F.}}:
\batitle{Nonlinear field equations for aligning self-propelled rods}.
\bjtitle{Phys. Rev. Lett.}
\bvolume{109},
\bfpage{268701}
(\byear{2012})
\end{barticle}
\endbibitem

%%% 14
\bibitem[\protect\citeauthoryear{Denk and Frey}{2020}]{Denk20}
\begin{barticle}
\bauthor{\bsnm{Denk}, \binits{J.}},
\bauthor{\bsnm{Frey}, \binits{E.}}:
\batitle{Pattern-induced local symmetry breaking in active-matter systems}.
\bjtitle{PNAS}
\bvolume{117},
\bfpage{31623}--\blpage{31630}
(\byear{2020})
\end{barticle}
\endbibitem

%%% 15
\bibitem[\protect\citeauthoryear{Pototsky et~al.}{2014}]{PTS14}
\begin{barticle}
\bauthor{\bsnm{Pototsky}, \binits{A.}},
\bauthor{\bsnm{Thiele}, \binits{U.}},
\bauthor{\bsnm{Stark}, \binits{H.}}:
\batitle{Stability of liquid films covered by a carpet of self-propelled
  surfactant particles}.
\bjtitle{Phys. Rev. E}
\bvolume{90},
\bfpage{030401}
(\byear{2014})
\end{barticle}
\endbibitem

%%% 16
\bibitem[\protect\citeauthoryear{Pototsky et~al.}{2016}]{PTS16}
\begin{barticle}
\bauthor{\bsnm{Pototsky}, \binits{A.}},
\bauthor{\bsnm{Thiele}, \binits{U.}},
\bauthor{\bsnm{Stark}, \binits{H.}}:
\batitle{Mode instabilities and dynamic patterns in a colony of self-propelled
  surfactant particles covering a thin liquid layer}.
\bjtitle{Eur. Phys. J E Soft Matter}
\bvolume{39},
\bfpage{51}
(\byear{2016})
\end{barticle}
\endbibitem

%%% 17
\bibitem[\protect\citeauthoryear{Oron et~al.}{1997}]{Oron97}
\begin{barticle}
\bauthor{\bsnm{Oron}, \binits{A.}},
\bauthor{\bsnm{Davis}, \binits{S.H.}},
\bauthor{\bsnm{Bankoff}, \binits{S.G.}}:
\batitle{Long-scale evolution of thin liquid films}.
\bjtitle{Rev. Mod. Phys.}
\bvolume{69},
\bfpage{931}--\blpage{980}
(\byear{1997})
\end{barticle}
\endbibitem

%%% 18
\bibitem[\protect\citeauthoryear{Merkt et~al.}{2005}]{MPBT2005pf}
\begin{barticle}
\bauthor{\bsnm{Merkt}, \binits{D.}},
\bauthor{\bsnm{Pototsky}, \binits{A.}},
\bauthor{\bsnm{Bestehorn}, \binits{M.}},
\bauthor{\bsnm{Thiele}, \binits{U.}}:
\batitle{Long-wave theory of bounded two-layer films with a free liquid-liquid
  interface: {S}hort- and long-time evolution}.
\bjtitle{Phys. Fluids}
\bvolume{17},
\bfpage{064104}
(\byear{2005})
\doiurl{10.1063/1.1935487}
\end{barticle}
\endbibitem

%%% 19
\bibitem[\protect\citeauthoryear{Frohoff-H{\"u}lsmann and
  Thiele}{2023}]{FrTh2023prl}
\begin{barticle}
\bauthor{\bsnm{Frohoff-H{\"u}lsmann}, \binits{T.}},
\bauthor{\bsnm{Thiele}, \binits{U.}}:
\batitle{Nonreciprocal {C}ahn-{H}illiard model emerges as a universal amplitude
  equation}.
\bjtitle{Phys. Rev. Lett.}
\bvolume{131},
\bfpage{107201}
(\byear{2023})
\doiurl{10.1103/PhysRevLett.131.107201}
\end{barticle}
\endbibitem

%%% 20
\bibitem[\protect\citeauthoryear{Marconi et~al.}{2008}]{MARCONI2008}
\begin{barticle}
\bauthor{\bsnm{Marconi}, \binits{U.M.B.}},
\bauthor{\bsnm{Puglisi}, \binits{A.}},
\bauthor{\bsnm{Rondoni}, \binits{L.}},
\bauthor{\bsnm{Vulpiani}, \binits{A.}}:
\batitle{Fluctuation–dissipation: Response theory in statistical physics}.
\bjtitle{Phys. Rep.}
\bvolume{461},
\bfpage{111}--\blpage{195}
(\byear{2008})
\end{barticle}
\endbibitem

%%% 21
\bibitem[\protect\citeauthoryear{Wu and Libchaber}{2000}]{Wu2000}
\begin{barticle}
\bauthor{\bsnm{Wu}, \binits{X.-L.}},
\bauthor{\bsnm{Libchaber}, \binits{A.}}:
\batitle{Particle diffusion in a quasi-two-dimensional bacterial bath}.
\bjtitle{Phys. Rev. Lett.}
\bvolume{84},
\bfpage{3017}--\blpage{3020}
(\byear{2000})
\end{barticle}
\endbibitem

%%% 22
\bibitem[\protect\citeauthoryear{Sokolov et~al.}{2007}]{Sokolov2007}
\begin{barticle}
\bauthor{\bsnm{Sokolov}, \binits{A.}},
\bauthor{\bsnm{Aranson}, \binits{I.S.}},
\bauthor{\bsnm{Kessler}, \binits{J.O.}},
\bauthor{\bsnm{Goldstein}, \binits{R.E.}}:
\batitle{Concentration dependence of the collective dynamics of swimming
  bacteria}.
\bjtitle{Phys. Rev. Lett.}
\bvolume{98},
\bfpage{158102}
(\byear{2007})
\end{barticle}
\endbibitem

%%% 23
\bibitem[\protect\citeauthoryear{Sokolov et~al.}{2010}]{Sokolov2010}
\begin{barticle}
\bauthor{\bsnm{Sokolov}, \binits{A.}},
\bauthor{\bsnm{Apodaca}, \binits{M.M.}},
\bauthor{\bsnm{Grzybowski}, \binits{B.A.}},
\bauthor{\bsnm{Aranson}, \binits{I.S.}}:
\batitle{Swimming bacteria power microscopic gears}.
\bjtitle{PNAS}
\bvolume{107}(\bissue{3}),
\bfpage{969}--\blpage{974}
(\byear{2010})
\end{barticle}
\endbibitem

%%% 24
\bibitem[\protect\citeauthoryear{Rongy and De~Wit}{2006}]{Rongy2006}
\begin{barticle}
\bauthor{\bsnm{Rongy}, \binits{L.}},
\bauthor{\bsnm{De~Wit}, \binits{A.}}:
\batitle{Steady {M}arangoni flow traveling with chemical fronts}.
\bjtitle{The Journal of Chemical Physics}
\bvolume{124},
\bfpage{164705}
(\byear{2006})
\end{barticle}
\endbibitem

%%% 25
\bibitem[\protect\citeauthoryear{Rongy and De~Wit}{2007}]{Rongy2007}
\begin{barticle}
\bauthor{\bsnm{Rongy}, \binits{L.}},
\bauthor{\bsnm{De~Wit}, \binits{A.}}:
\batitle{Marangoni flow around chemical fronts traveling in thin solution
  layers: influence of the liquid depth}.
\bjtitle{J. Eng. Math.}
\bvolume{59},
\bfpage{1573}--\blpage{2703}
(\byear{2007})
\end{barticle}
\endbibitem


%%% 26
\bibitem[\protect\citeauthoryear{Anderson}{1989}]{Anderson89}
\begin{barticle}
\bauthor{\bsnm{Anderson}, \binits{J.L.}}:
\batitle{Colloid transport by interfacial forces}.
\bjtitle{Annu. Rev. Fluid Mech.}
\bvolume{21},
\bfpage{61}--\blpage{99}
(\byear{1989})
\end{barticle}
\endbibitem

%%% 27
\bibitem[\protect\citeauthoryear{Feng et~al.}{2023}]{Feng23}
\begin{barticle}
\bauthor{\bsnm{Feng}, \binits{C.}},
\bauthor{\bsnm{Molina}, \binits{J.J.}},
\bauthor{\bsnm{Turner}, \binits{M.S.}},
\bauthor{\bsnm{Yamamoto}, \binits{R.}}:
\batitle{Dynamics of microswimmers near a liquid–liquid interface with
  viscosity difference}.
\bjtitle{Phys, Fluids}
\bvolume{35}(\bissue{5}),
\bfpage{051903}
(\byear{2023})
\end{barticle}
\endbibitem

%%% 28
\bibitem[\protect\citeauthoryear{Bertin et~al.}{2006}]{Bertin2006}
\begin{barticle}
\bauthor{\bsnm{Bertin}, \binits{E.}},
\bauthor{\bsnm{Droz}, \binits{M.}},
\bauthor{\bsnm{Gr\'egoire}, \binits{G.}}:
\batitle{Boltzmann and hydrodynamic description for self-propelled particles}.
\bjtitle{Phys. Rev. E}
\bvolume{74},
\bfpage{022101}
(\byear{2006})
\end{barticle}
\endbibitem

%%% 29
\bibitem[\protect\citeauthoryear{Thiele et~al.}{2012}]{ThAP2012pf}
\begin{barticle}
\bauthor{\bsnm{Thiele}, \binits{U.}},
\bauthor{\bsnm{Archer}, \binits{A.J.}},
\bauthor{\bsnm{Plapp}, \binits{M.}}:
\batitle{Thermodynamically consistent description of the hydrodynamics of free
  surfaces covered by insoluble surfactants of high concentration}.
\bjtitle{Phys. Fluids}
\bvolume{24},
\bfpage{102107}
(\byear{2012})
\doiurl{10.1063/1.4758476}
\end{barticle}
\endbibitem

%%% 30
\bibitem[\protect\citeauthoryear{Thess et~al.}{1997}]{Thess97}
\begin{barticle}
\bauthor{\bsnm{Thess}, \binits{A.}},
\bauthor{\bsnm{Spirn}, \binits{D.}},
\bauthor{\bsnm{J{\"u}ttner}, \binits{B.}}:
\batitle{A two-dimensional model for slow convection at infinite {M}arangoni
  number}.
\bjtitle{J. Fluid Mech.}
\bvolume{331},
\bfpage{283}--\blpage{312}
(\byear{1997})
\end{barticle}
\endbibitem

%%% 31
\bibitem[\protect\citeauthoryear{Thiele}{2007}]{Thie2007chapter}
\begin{bchapter}
\bauthor{\bsnm{Thiele}, \binits{U.}}:
\bctitle{Structure formation in thin liquid films}.
In: \beditor{\bsnm{Kalliadasis}, \binits{S.}},
\beditor{\bsnm{Thiele}, \binits{U.}} (eds.)
\bbtitle{Thin Films of Soft Matter},
pp. \bfpage{25}--\blpage{93}.
\bpublisher{Springer},
\blocation{Vienna}
(\byear{2007}).
\doiurl{10.1007/978-3-211-69808-2\_2}
\end{bchapter}
\endbibitem

%%% 32
\bibitem[\protect\citeauthoryear{Craster and Matar}{2009}]{CrMa2009rmp}
\begin{barticle}
\bauthor{\bsnm{Craster}, \binits{R.V.}},
\bauthor{\bsnm{Matar}, \binits{O.K.}}:
\batitle{Dynamics and stability of thin liquid films}.
\bjtitle{Rev. Mod. Phys.}
\bvolume{81},
\bfpage{1131}--\blpage{1198}
(\byear{2009})
\doiurl{10.1103/RevModPhys.81.1131}
\end{barticle}
\endbibitem

%%% 33
\bibitem[\protect\citeauthoryear{}{}]{note}
\begin{botherref}
We find similar wave dynamics when the flow is given by Eqs.\,(\ref{eq2}).
\end{botherref}
\endbibitem

%%% 34
\bibitem[\protect\citeauthoryear{Glasner et~al.}{2009}]{GORS2009ejam}
\begin{barticle}
\bauthor{\bsnm{Glasner}, \binits{K.}},
\bauthor{\bsnm{Otto}, \binits{F.}},
\bauthor{\bsnm{Rump}, \binits{T.}},
\bauthor{\bsnm{Slepcev}, \binits{D.}}:
\batitle{Ostwald ripening of droplets: the role of migration}.
\bjtitle{Eur. J. Appl. Math.}
\bvolume{20},
\bfpage{1}--\blpage{67}
(\byear{2009})
\doiurl{10.1017/S0956792508007559}
\end{barticle}
\endbibitem

%%% 35
\bibitem[\protect\citeauthoryear{Pototsky et~al.}{2014}]{PTA14}
\begin{barticle}
\bauthor{\bsnm{Pototsky}, \binits{A.}},
\bauthor{\bsnm{Thiele}, \binits{U.}},
\bauthor{\bsnm{Archer}, \binits{A.J.}}:
\batitle{Coarsening modes of clusters of aggregating particles}.
\bjtitle{Phys. Rev. E}
\bvolume{89},
\bfpage{032144}
(\byear{2014})
\end{barticle}
\endbibitem

%%% 36
\bibitem[\protect\citeauthoryear{Zaks et~al.}{2005}]{Zaks05}
\begin{barticle}
\bauthor{\bsnm{Zaks}, \binits{M.A.}},
\bauthor{\bsnm{Podolny}, \binits{A.}},
\bauthor{\bsnm{Nepomnyashchy}, \binits{A.A.}},
\bauthor{\bsnm{Golovin}, \binits{A.A.}}:
\batitle{Periodic stationary patterns governed by a convective cahn--hilliard
  equation}.
\bjtitle{SIAP}
\bvolume{66},
\bfpage{700}--\blpage{720}
(\byear{2005})
\end{barticle}
\endbibitem

%%% 37
\bibitem[\protect\citeauthoryear{Tseluiko et~al.}{2020}]{TALT2020n}
\begin{barticle}
\bauthor{\bsnm{Tseluiko}, \binits{D.}},
\bauthor{\bsnm{Alesemi}, \binits{M.}},
\bauthor{\bsnm{Lin}, \binits{T.-S.}},
\bauthor{\bsnm{Thiele}, \binits{U.}}:
\batitle{Effect of driving on coarsening dynamics in phase-separating systems}.
\bjtitle{Nonlinearity}
\bvolume{33},
\bfpage{4449}--\blpage{4483}
(\byear{2020})
\doiurl{10.1088/1361-6544/ab8bb0}
%{\href{https://arxiv.org/abs/http://arxiv.org/abs/1905.13396}{{http://arxiv.org/abs/1905.13396}}}
\end{barticle}
\endbibitem

%%% 38
\bibitem[\protect\citeauthoryear{Frohoff-H\"ulsmann et~al.}{2021}]{Frohoff2020}
\begin{barticle}
\bauthor{\bsnm{Frohoff-H\"ulsmann}, \binits{T.}},
\bauthor{\bsnm{Wrembel}, \binits{J.}},
\bauthor{\bsnm{Thiele}, \binits{U.}}:
\batitle{Suppression of coarsening and emergence of oscillatory behavior in a
  Cahn-Hilliard model with nonvariational coupling}.
\bjtitle{Phys. Rev. E}
\bvolume{103},
\bfpage{042602}
(\byear{2021})
\end{barticle}
\endbibitem

%%% 39
\bibitem[\protect\citeauthoryear{Palacci et~al.}{2010}]{Palacci2010}
\begin{barticle}
\bauthor{\bsnm{Palacci}, \binits{J.}},
\bauthor{\bsnm{Ab\'ecassis}, \binits{B.}},
\bauthor{\bsnm{Cottin-Bizonne}, \binits{C.}},
\bauthor{\bsnm{Ybert}, \binits{C.}},
\bauthor{\bsnm{Bocquet}, \binits{L.}}:
\batitle{Colloidal motility and pattern formation under rectified
  diffusiophoresis}.
\bjtitle{Phys. Rev. Lett.}
\bvolume{104},
\bfpage{138302}
(\byear{2010})
\end{barticle}
\endbibitem

%%% 40
\bibitem[\protect\citeauthoryear{Palacci et~al.}{2014}]{Palacci2014}
\begin{barticle}
\bauthor{\bsnm{Palacci}, \binits{J.}},
\bauthor{\bsnm{Sacanna}, \binits{S.}},
\bauthor{\bsnm{Kim}, \binits{S.-H.}},
\bauthor{\bsnm{Yi}, \binits{G.-R.}},
\bauthor{\bsnm{Pine}, \binits{D.J.}},
\bauthor{\bsnm{Chaikin}, \binits{P.M.}}:
\batitle{Light-activated self-propelled colloids}.
\bjtitle{Philos. Trans. R. Soc. A: Mathematical, Physical and Engineering
  Sciences}
\bvolume{372},
\bfpage{20130372}
(\byear{2014})
\end{barticle}
\endbibitem

%%% 41
\bibitem[\protect\citeauthoryear{Palacci et~al.}{2013}]{Palacci2020}
\begin{barticle}
\bauthor{\bsnm{Palacci}, \binits{J.}},
\bauthor{\bsnm{Sacanna}, \binits{S.}},
\bauthor{\bsnm{Steinberg}, \binits{A.P.}},
\bauthor{\bsnm{Pine}, \binits{D.J.}},
\bauthor{\bsnm{Chaikin}, \binits{P.M.}}:
\batitle{Living crystals of light-activated colloidal surfers}.
\bjtitle{Science}
\bvolume{339},
\bfpage{936}--\blpage{940}
(\byear{2013})
\doiurl{10.1126/science.1230020}
%{\href{https://arxiv.org/abs/https://www.science.org/doi/pdf/10.1126/science.1230020}{{https://www.science.org/doi/pdf/10.1126/science.1230020}}}
\end{barticle}
\endbibitem

%%% 42
\bibitem[\protect\citeauthoryear{Gonnella et~al.}{2015}]{GONNELLA2015}
\begin{barticle}
\bauthor{\bsnm{Gonnella}, \binits{G.}},
\bauthor{\bsnm{Marenduzzo}, \binits{D.}},
\bauthor{\bsnm{Suma}, \binits{A.}},
\bauthor{\bsnm{Tiribocchi}, \binits{A.}}:
\batitle{Motility-induced phase separation and coarsening in active matter}.
\bjtitle{Comptes Rendus Physique}
\bvolume{16},
\bfpage{316}--\blpage{331}
(\byear{2015}).
%\bcomment{Coarsening dynamics / Dynamique de coarsening}
\end{barticle}
\endbibitem

%%% 43
\bibitem[\protect\citeauthoryear{Greve and Thiele}{2024}]{GrTh2024c}
\begin{barticle}
\bauthor{\bsnm{Greve}, \binits{D.}},
\bauthor{\bsnm{Thiele}, \binits{U.}}:
\batitle{An amplitude equation for the conserved-{H}opf bifurcation --
  derivation, analysis and assessment}.
\bjtitle{Chaos}
\bvolume{34},
\bfpage{123134}
(\byear{2024})
\doiurl{10.1063/5.0222013}
\end{barticle}
\endbibitem

%%% 44
\bibitem[\protect\citeauthoryear{Krauskopf
  et~al.}{2007}]{KrauskopfOsingaGalan-Vioque2007}
\begin{bbook}
\beditor{\bsnm{Krauskopf}, \binits{B.}},
\beditor{\bsnm{Osinga}, \binits{H.M.}},
\beditor{\bsnm{Galan-Vioque}, \binits{J.}} (eds.):
\bbtitle{Numerical Continuation Methods for Dynamical Systems}.
\bpublisher{Springer},
\blocation{Dordrecht}
(\byear{2007}).
\doiurl{10.1007/978-1-4020-6356-5}
\end{bbook}
\endbibitem

%%% 45
\bibitem[\protect\citeauthoryear{Engelnkemper et~al.}{2019}]{EGUW2019springer}
\begin{bchapter}
\bauthor{\bsnm{Engelnkemper}, \binits{S.}},
\bauthor{\bsnm{Gurevich}, \binits{S.V.}},
\bauthor{\bsnm{Uecker}, \binits{H.}},
\bauthor{\bsnm{Wetzel}, \binits{D.}},
\bauthor{\bsnm{Thiele}, \binits{U.}}:
\bctitle{Continuation for thin film hydrodynamics and related scalar problems}.
In: \beditor{\bsnm{Gelfgat}, \binits{A.}} (ed.)
\bbtitle{Computational Modeling of Bifurcations and Instabilities in Fluid
  Mechanics}.
\bsertitle{Computational Methods in Applied Sciences, vol 50},
pp. \bfpage{459}--\blpage{501}.
\bpublisher{Springer},
\blocation{Cham}
(\byear{2019}).
\doiurl{10.1007/978-3-319-91494-7_13}
\end{bchapter}
\endbibitem

%%% 46
\bibitem[\protect\citeauthoryear{Doedel and Oldeman}{2009}]{DoedelOldeman2009}
\begin{bbook}
\bauthor{\bsnm{Doedel}, \binits{E.J.}},
\bauthor{\bsnm{Oldeman}, \binits{B.E.}}:
\bbtitle{AUTO07p: Continuation and Bifurcation Software for Ordinary
  Differential Equations}.
\bpublisher{Concordia University},
\blocation{Montreal}
(\byear{2009}).
\burl{https://www.macs.hw.ac.uk/\texttildelow gabriel/auto07/auto.html}
\end{bbook}
\endbibitem

\end{thebibliography}

\end{document}